\documentclass[showpacs,twocolumn,floatfix,amsmath,amssymb,aps,prb]{revtex4}
\usepackage{graphicx}
\usepackage{latexsym,epsfig,bm,times,psfrag,subfigure} 
\usepackage{bm,times} 
\usepackage{dcolumn}% Align table columns on decimal point
\begin{document}

%#######
\newcommand{\nn}{m}
\newcommand{\CT}{\mathcal{T}}
\newcommand{\CR}{\mathcal{R}}
\newcommand{\rt}{r}
\newcommand{\rtd}{t}
\newcommand{\vL}{\vF}
\newcommand{\dagg}{^{\dagger}}
\newcommand{\phdagg}{^{\phantom{\dagger}}}
\renewcommand{\Im}{\textrm{Im}}
\renewcommand{\Re}{\textrm{Re}}
\newcommand{\vF}{v_{\textrm{F}}}
\newcommand{\vO}{v_{\textrm{0}}}
\newcommand{\kB}{k_{\textrm{B}}}
 \newcommand{\sig}{G_{0}}
\newcommand{\fluxphase}{\theta}
\newcommand{\A}{{\cal A}}
\newcommand{\n}{{\bf n}}
\newcommand{\m}{{\bf m}}
\newcommand{\Tr}{\textrm{Tr}}
\newcommand{\Ham}{\mathcal{H}}
\newcommand{\order}[1]{\mathcal{O}(#1)}
\newcommand{\ket}[1]{\vert #1\rangle}
\newcommand{\bra}[1]{\langle #1\vert}
\newcommand{\mb}[1]{\mathbf{#1}}

%#####################################################
\title{Decoherence and interactions in an electronic Mach-Zehnder interferometer} 
\author{J.\ T.\ Chalker,$^{1}$ Yuval\
Gefen,$^{1,2}$ and M. Y. Veillette$^1$}
\affiliation{$^{1}$Theoretical Physics, Oxford University, 1, Keble Road, Oxford, OX1 3NP, United Kingdom\\}
\affiliation{$^{2}$Department of Condensed Matter Physics, Weizmann Institute
of Science, Rehevot, 76100 Israel\\}
\date{\today} 
%############################################################
\begin{abstract}
We develop a theoretical description of a Mach-Zehnder interferometer
built from integer quantum Hall edge states, with an emphasis on how
electron-electron interactions produce decoherence. We calculate
the visibility of interference fringes and noise power, as a function
of bias voltage and of temperature. Interactions are treated exactly,
by using bosonization and considering edge states that are only weakly
coupled via tunneling at the interferometer beam-splitters. In this
weak-tunneling limit, we show that the bias-dependence of Aharonov-Bohm
oscillations in source-drain conductance and noise power provides a
direct measure of the one-electron correlation function for an isolated quantum
Hall edge state. We find the asymptotic form of this correlation function
for systems with either short-range interactions or unscreened Coulomb
interactions, extracting a dephasing length $\ell_{\varphi}$ that
varies with temperature $T$ as $\ell_{\varphi} \propto T^{-3}$ in the
first case and as $\ell_{\varphi} \propto T^{-1} \ln^2(T)$ in the second case.
\end{abstract}
%############################################################
\maketitle

\section{Introduction}

Several striking phenomena in electronic Mach-Zehnder
interferometers built from integer quantum Hall edge states have been
reported in a sequence of recent experimental
papers.\cite{heiblum1,heiblum2,heiblum3,recent-expt}
The central observation\cite{heiblum1} is of 
interference fringes in the differential source-drain conductance of the
interferometer, as the magnetic flux enclosed between its
arms is changed. The high contrast of these fringes (up to 60\%)
is remarkable in view of the relatively large size (around 10$\mu{\rm
  m}$) of the interferometer. The interference fringes are suppressed
as sample temperature is increased, or as the system is driven out of
equilibrium by finite source-drain bias. In each case, this
suppression offers a window onto
dephasing processes in the system, and such processes are the
subject of this paper.

These experiments form part of a larger effort to understand and make
use of interference effects in quantum Hall systems. In the context of
the fractional quantum Hall effect, challenges include detection
of anyonic phases for
quasiparticles\cite{goldman,Jonckheere,Law,Kim,nagger,averin,stern} 
and, more ambitiously, of non-abelian
quasiparticle phases.\cite{stern} Against this background, it is clearly
important to have a proper understanding of the simpler interaction
effects expected in integer quantum Hall systems, and our aim
in the following is to establish a theoretical description
of some of these effects.

\begin{figure}[ht]
\begin{center}
\includegraphics[width=7cm]{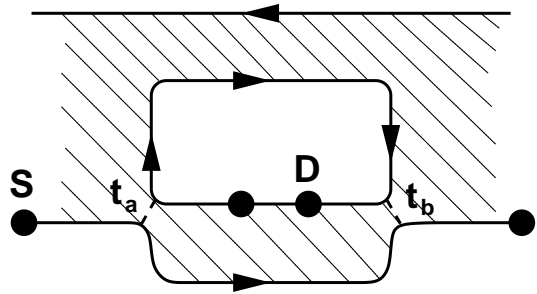}
\caption{Sketch of the electronic Mach-Zehnder interferometer. Two quantum Hall edge
  states are coupled by tunneling at two quantum point contacts. One
  of these edge states lies at the edge of a
  Hall bar and starts from source $S$; the other encircles an antidot
  within the Hall bar and ends at drain $D$.
  Tunneling amplitudes at
  the point contacts are denoted by $t_a$ and $t_b$.
}
 \label{mz-expt}
\end{center}
\end{figure}

The design of an electronic Mach-Zehnder interferometer is illustrated in Fig.~\ref{mz-expt}:
two edge states, which propagate in the {\it same} sense, meet at two places where
they are coupled by quantum point contacts, which
act as the equivalent of the beam-splitters used in the optical
version of the interferometer.
An attractive feature of the Mach-Zehnder arrangement is that each electron passes only
once through the interferometer, and
this simplifies the theoretical analysis.
In contrast, for a Fabry-P\'erot design,\cite{chamon} particles may execute
multiple circuits of the interferometer before exiting.
%%%%%%%%%%%%%%%%%%%%%%%%%%%[3]%%%%%%%%%%%%%%%%%%%%%%%%%%%%%%%%%%%%%%%%%%%%%%%
%%%%%%%%%%%%%%%%%%%%%%%%%%%%%%%%%%%%%%%%%%%%%%%%%%%%%%%%%%%%%%%%%%%%%%%%

A body of earlier theoretical
work\cite{Seelig01,Marquardt04,Forster,Chung,Marquardt06,neder-marquardt}
on electronic Mach-Zehnder
interferometers built from integer quantum Hall edge states has
treated sources of dephasing that are distinct
from the edge states themselves. These sources of dephasing are
represented by a fictitious voltage probe\cite{Chung}
or by a fluctuating electrostatic 
potential,\cite{Seelig01,Marquardt04,Forster,Marquardt06,neder-marquardt} which
may have various origins, including the movement of charged impurities
within the sample, and thermal or non-equilibrium electromagnetic
radiation in the sample's surroundings. Our concern in this paper is
instead with dephasing that is {\it intrinsic} to the interferometer,
and that arises from interactions between electrons in the edge states
themselves. 
Some previous discussions \cite{heiblum1,Marquardt04} have emphasised
the distinction between {\it fast} and {\it slow} potential 
fluctuations, as compared to the flight time for electrons passing
through the interferometer. Within that classification, we deal here 
with fast fluctuations.
We are motivated by several considerations. First, as we
show, intrinsic dephasing is a useful
experimental probe of electron-electron interactions in integer quantum Hall
edge states. Second, since
external sources of dephasing can, at least in principle, be reduced in experiment,
intrinsic contributions should ultimately dominate. And third, in
relation to
recent experiments, while temperature-dependent dephasing may
plausibly arise from either external or instrinsic sources, we
believe that bias-dependent dephasing most likely results from an
intrinsic mechanism.

\begin{figure}[ht]
\begin{center}
\includegraphics[width=7cm]{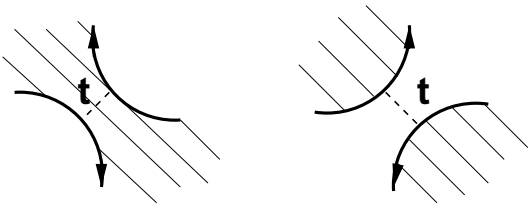}
\caption{Action of the tunneling Hamiltonian, for a contact that is
  almost open (left), and for one that is almost pinched-off (right);
the shaded region is occupied by electrons. 
}
 \label{point-contact}
\end{center}
\end{figure}

Our approach is to start from a system without tunneling at the
interferometer quantum point contacts. In this limit, assuming a
confining potential for electrons at the sample edge that rises linearly
with position, electron-electron interactions can be handled exactly,
using bosonization to obtain a harmonic Hamiltonian for collective
edge excitations. For simplicity, we study translation-invariant edges
and omit interactions between electrons on different
edges. Introducing tunneling into the description, since 
the tunneling Hamiltonian is not quadratic in boson operators, we treat it
perturbatively, calculating source-drain conductance and noise power
for the interferometer at leading order in tunneling amplitudes.
Such an approach is good both when point contacts are almost open
and the reflection probability is small, as
illustrated in Fig.~\ref{mz-expt}, and when they are close to pinch-off 
and the transmission probability is small; for the two cases, the 
tunneling Hamiltonian acts as sketched in Fig.~\ref{point-contact}.
In this way, we bracket
from both sides the regime in which transmission and reflection
probabilities are similar.

It is a familiar fact that interactions have rather limited
consequences in integer quantum Hall edge states, in the sense that
the asymptotic low-energy behaviour is that of a Fermi liquid rather than
a Luttinger liquid.\cite{wen} Indeed, if interactions are modelled by
a short-range potential, their only effect is to renormalise the
edge-state velocity. More generally, interactions generate a
non-linear dispersion relation for collective modes. The dispersion,
in turn, means that a wavepacket representing an electron that has
tunneled into an edge will spread as it propagates. It is this spreading
that gives rise to the dephasing we study here. In the context of an
interferometer, we are concerned with what happens during
propagation along an interferometer arm. If the arm-length is large,
the propagation time is long, and spreading is determined by the low-frequency form
of the dispersion relation, which depends on the behaviour of the
Fourier transform of the interaction potential at small wavevector.
As an intermediate step in the calculation of observable quantities,
we find the asymptotic, long-distance form of the one-electron
correlation function,
presenting results for both generic, finite-range interactions and for
unscreened Coulomb interactions. From this we obtain the
bias-dependence of conductance and noise power for the interferometer.
Our approach is parallel to recent calculations of the effect of
Coulomb interactions on transport between edge states in multilayer
quantum Hall systems.\cite{Tomlinson} We do not consider
inter-edge interactions in special geometries, which may give rise to
resonance phenomena as 
discussed recently in Ref.~\onlinecite{cheianov}.

Without interactions, Aharonov-Bohm oscillations in conductance
contain only the zeroth and first harmonics in $\Phi/\Phi_0$, while
those of the noise power contain zeroth, first and second harmonics.
From the structure of our calculations, it is apparent that
interactions generate higher harmonics for both quantities, albeit
at higher orders in our expansion in tunneling amplitudes.

The remainder of the paper is organised as follows. In
Sec.~\ref{model} we define our model and the one-electron correlation function for
an isolated edge, from which interferometer properties at weak
tunneling can be calculated. We derive expressions for conductance and
noise power in Sec.~\ref{sec:cond}. We discuss asymptotics
of the correlation function in Sec.~\ref{sec:correlation-fn} and use these in
Sec.~\ref{sec:visibility} to determine visibility of interference fringes in
conductance as a function of bias voltage and temperature. We discuss
our findings in Sec.~\ref{discussion}.

\section{Model}\label{model}

\begin{figure}[ht]
\begin{center}
\includegraphics[width=7cm]{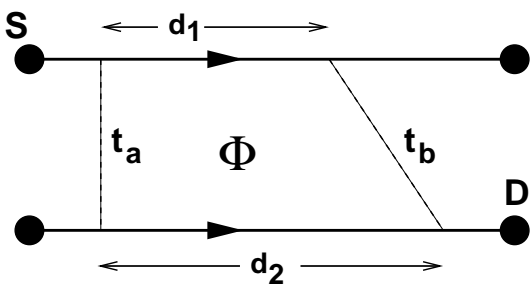}
\caption{Schematic view of interferometer, showing
arm lengths $d_1$ and $d_2$, tunneling amplitudes $t_a$ and $t_b$, and
arrangement of source $S$ and drain $D$. The magnetic flux 
enclosed by the two possible electron paths is indicated by
$\Phi$.}
 \label{mz}
\end{center}
\end{figure}

We consider the idealised version of the experimental system shown
in Fig.~\ref{mz}, with the model 
Hamiltonian $\Ham = \Ham_0 + \Ham_{\rm tun} + \Ham_{\rm int}$.
Here, the single-particle terms $\Ham_0$ and $\Ham_{\textrm{tun}}$
represent free motion along each edge and
inter-edge electron tunneling, respectively, while
$\Ham_{\textrm{int}}$ describes electron interactions within an
edge.  We write the Hamiltonian in terms of the electron creation
operator $c\dagg_{qm}$  for a state with wavevector $q$ on 
the edge $\nn =\{ 1,2 \}$, defining one-particle basis states for an
edge of length $L$ so that $q=2\pi n_q/L$, where $n_q$ is integer.
The creation operator at a point $x$ on the edge $\nn$ is
\begin{equation}
\psi_{\nn}^{\dagger}(x)=\frac{1}{\sqrt{L}}\sum_{q=-\infty}^{\infty}
e^{-i q x }c_{q \nn}^{\dagger}
\label{eq:ctopsi}\,.
\end{equation}
We normal order the Hamiltonian with respect to a vacuum in which
states are occupied for $q\leq 0$ and empty otherwise.  Taking a
strictly linear electron dispersion relation, the Hamiltonian for free
motion is
\begin{eqnarray}
\Ham_0&=& -i\hbar v \sum_{\nn=1,2} \int \! \!  dx \:
:\!\psi^{\dagger}_\nn(x)\partial_x\psi_\nn(x)\!:
\label{eq:H0psi}\,
\end{eqnarray}
where $v$ is the edge velocity; the consequences of curvature
in this dispersion relation  are discussed in Sec.~\ref{sec:correlation-fn}.
We consider narrow quantum point contacts, so
that the tunneling occurs at a single position, with 
\begin{eqnarray}
\Ham_{\textrm{tun}}&= t_{a}\psi^{\dagger}_{1}(0)
\psi_2(0)+ t_{b}\psi^{\dagger}_{1}(d_{1})
\psi_2(d_{2})+\textrm{H.~c.}\, .
\label{1}
\end{eqnarray}
The electron-electron interaction energy within each edge can be written in 
terms of the density operator $\rho_\nn(x)=\psi^{\dagger}_\nn(x)
\psi_\nn(x)$. Assuming a translation-invariant 
two-particle potential $U(x-x')$ we have
\begin{equation}
\Ham_{\textrm{int}} =\frac{1}{2}\sum_{\nn=1,2} \iint  \! \!  dx \:
\:dx^{\prime}\: :\rho_{\nn}(x) U(x-x^{\prime}) \rho_{\nn}(x^{\prime}):\,.
\label{2}
\end{equation}
The phase arising from the enclosed flux $\Phi$ 
appears in the tunneling amplitudes: $t_a t_b^* = |t_a
t_b^*|e^{i\theta}$
where $\theta = \Phi/\Phi_0$ and $\Phi_0={h}/{e}$ is the flux quantum.
%%%%%%%%%%%%%%%%%%%%%%%%%%%[3]%%%%%%%%%%%%%%%%%%%%%%%%%%%%%%%%%%%%%%%%%%%%%%%
%%%%%%%%%%%%%%%%%%%%%%%%%%%%%%%%%%%%%%%%%%%%%%%%%%%%%%%%%%%%%%%%%%%%%%%%

\subsection{Bosonized Hamiltonian}\label{ssec:model:bose}

We bosonize the Hamiltonian in the standard way, expressing
$\Ham_{\textrm{0}}+\Ham_{\textrm{int}}$ in terms of harmonic
collective modes. 
Boson creation operators are defined (see, for
example, Ref.~\onlinecite{vonDelft}) as
\begin{equation}
b_{q m}^{\dagger}=\frac{i}{\sqrt{n_q}}\sum_{k}
c^{\dagger}_{k+q,m} c\phdagg_{k,m}
\label{3}
\end{equation}
for $q>0$. 
These boson operators satisfy the canonical commutation relation 
\begin{equation}
\label{CCR}
\left[ b^{\phantom{\dagger}}_{q m}, b^{\dagger}_{k n} \right]= 
\delta_{q k} \delta_{m n}\,.
\end{equation}
Fourier transforming the interaction potential and expressing the result as a
velocity, we introduce
\begin{equation}
u(q) =\frac{1}{2\pi \hbar}\int \! \!  dx \: e^{i q x}U(x)\,.
\label{4}
\end{equation}
%The Fermi velocity renormalized by Hartree interactions is
%$\vF = \vO -  u(0)$, where the divergence 
%which arises in the sum in the case of Coulomb interactions
%is cancelled by contributions to $v$ from a neutralizing background.
The Hamiltonian in the absence of tunneling is $\Ham_1 \equiv \Ham_0 +
\Ham_{\textrm{int}}$, with
\begin{equation}
\Ham_1=\sum_{\nn=1,2}\left[ \frac{\pi \hbar v}{L}{\hat
N_{\nn}}({\hat N_{\nn}}+1) +\sum_{q>0} \hbar \omega(q)
b\dagg_{q \nn}b\phdagg_{q \nn}\ \right]\,,
\label{5}
\end{equation}
where the collective mode frequency is $\omega(q){=}q \left[v + u(q) - u(0)\right]$,
and the number operator for edge ${\nn}$ is ${\hat N}_{\nn}{=} \int
\!   dx \:
\rho_{\nn}(x) $.
Linearising the dependence on ${\hat N}_m$ about
the mean value $\langle {\hat N_{\nn}} \rangle$, 
expressed in terms of the chemical potential via
$\mu_{\nn}= 2 \pi \hbar v \langle {\hat N_{\nn}} \rangle/L $,
and omitting a constant, we have

\begin{equation}
\Ham_1=\sum_{\nn=1,2} \left[ \mu_{\nn} {\hat N_{\nn}}+ \sum_{q>0} \hbar \omega(q)
b\dagg_{q \nn}b\phdagg_{q \nn} \right]\,.
\label{6}
\end{equation}

We will focus on the limit in which the interferometer
arm lengths $d_1$ and $d_2$ are much larger than the range of
interactions, and will therefore be concerned with 
the dispersion relation at small $\omega(q)$, and hence small $q$.
We consider two alternative forms for interactions: generic
short-range interactions, and unscreened Coulomb interactions.
The first is appropriate for an edge state defined by a
metallic gate, since in this case charges in the gate will screen
interactions between electrons in the edge state. The second case
is appropriate for an edge state defined by etching.

For short-range
interactions we write
$u(0)- u(q) = v (b q)^{n-1} +{\cal O}(q^n)$, where the length
$b$ characterises a combination of the range and strength
%(in units related to $v$)
of interactions. Since the
Fourier transform of a short-range interaction potential
is analytic at $q=0$, one expects $n=3$. In this way, an interaction
$U(x)$ of strength $U_0$ and range $b_0$ is described completely
for our purposes by the length $b \sim b_0 \cdot (U_0 b_0 /
\hbar v)^{1/2} $.

For the Coulomb interaction, regularized at short distances by a 
finite width $a$ of edge states, with the form
\begin{equation}
U(x)=\frac{e^2}{4\pi\epsilon_0\epsilon_r}\frac{1}{\sqrt{x^2 + a^2}}
\label{eq:rsCoulomb}\,
\end{equation}
one has
\begin{equation}
u(q)= u K_0 ( a q) \, ,
\label{7}
\end{equation}
where $K_0$ is the hyperbolic Bessel function and
$u={e^2}/{(2 \pi)^2 \hbar \epsilon_0\epsilon_r}$. 
At small
$q$ one has in this case 
$\omega(q) = vq+uq\ln([aq]^{-1})$.

We emphasise that in our treatment of long-wavelength edge excitations, all interaction effects appear in
the dispersion relation $\omega(q)$ for collective modes via the
potential $u(q)$. The bare edge velocity $v$ is determined solely
by the gradient of the external potential that confines electrons
within the system, which does not incorporate screening effects
arising from the two-dimensional electron gas. 
This is the case even though it is expected, for realistic parameters,
that the edges of a quantum Hall system are compressible 
and that screening by electrons within the edge is important, as
represented by a Hartree potential that varies rapidly
across the edge.\cite{Chklovskii} This Hartree contribution to the
confining potential does not influence the velocity of modes
with wavelength much longer than the width of a compressible edge,
because when such modes are excited, the entire charge distribution within
the compressible edge is displaced rigidly, carrying the
Hartree potential with it.\cite{han-thouless}

\subsection{Correlation function}\label{ssec:model:G}

A central quantity in our calculation of transport properties is
the one-particle 
correlation function in the absence of tunneling. Note that, without
tunneling, 
%$\Ham=\Ham_{0}+\Ham_{\textrm{int}}$ and 
the two edges may be in
equilibrium with two separate reservoirs at different chemical potentials.
We use the interaction representation, writing
$\mathcal{O}(t)=e^{i\Ham_1 t/\hbar}\mathcal{O}e^{-i\Ham_1 t/\hbar}$.
%where $\Ham_1 = \Ham_0 + \Ham_{\rm int}$.
Thermal averages in the absence of tunneling are denoted by
$\langle\ldots\rangle\equiv\textrm{Tr}(e^{-\beta
\Ham_1}\ldots)/\textrm{Tr}(e^{-\beta \Ham_1})$.
The one-electron correlation function we are concerned with is
\begin{equation}
G_{\nn}(x,t)\equiv\langle\psi_{\nn}^{\dagger}(x,t)\psi
\phdagg_{\nn}(0,0)\rangle\label{eq:defG}\,.
\end{equation}

We evaluate this in a standard way, using bosonization.
As a first step, define the boson field operator 
\begin{equation}
\phi_{\nn}(x)=-\sum_{q>0} \frac{1}{\sqrt{n_q}} \:
\left(e^{-iqx}b_{q, \nn}^{\dagger}+ e^{iqx}b_{q, \nn}\right)e^{-\epsilon q/2}
\label{eq:defphiboson}
\end{equation}
where $\epsilon$ is an ultraviolet cut-off.
This field obeys the commutation relation
\begin{equation}
[\phi_{n}(x), \partial_y\phi_{m}(y) ] = -2 \pi i 
\delta_{n m} \delta(x-y)\,.
\end{equation}
The fermion and boson field operators are related by
\begin{equation}
\psi_{\nn}(x)=\frac{F_{\nn}}{ \sqrt{2\pi\epsilon}}  e^{i 2\pi {\hat N}_{\nn}
x/L} e^{-i\phi_{\nn}(x)}\, 
\label{22}
\end{equation}
where $F_{\nn}$ is a Klein factor that satisfies the 
anticommutation relation $\{ F^{\dagger}_{\nn}, F^{\phantom{\dagger}}_{n} \}
= 2 \delta_{\nn n}$. 
The correlation function can be written as
\begin{eqnarray}
G_{\nn}(x,t)\!\!=&\!\frac{1}{2\pi\epsilon}e^{-i \mu_{\nn} x/(\hbar
  v)}
\langle
F^{\dagger}_{\nn}(t) F_{\nn}(0) \rangle
\nonumber\\
&\times \langle e^{i\phi_{\nn}(x,t)}e^{-i\phi_{\nn}(0,0)}\rangle\,.
\label{23}
\end{eqnarray}
The two expectation values are straightforward to evaluate. 
The one involving Klein factors yields
\begin{equation}
\langle F^{\dagger}_{\nn}(t) F_{\nn}(0) \rangle = e^{i \mu_{\nn} t / \hbar}\,.
 \label{24}
\end{equation}
To calculate the bosonic expectation value, it is useful
to define the function $g_{\nn}(r,t)$ and its logarithm $S_{\nn}(r,t)$ via
\begin{equation}
g_{\nn}(x,t)
\equiv
\frac{1}{2 \pi \epsilon} \langle
e^{i\phi_{\nn}(x,t)}e^{-i\phi_{\nn}(0,0)}\rangle 
\equiv
\frac{1}{(2\pi)}e^{S_{\nn}(x,t)} 
\label{eq:GtoS}\,.
\end{equation}
Since the bosonic Hamiltonian is quadratic, we can express $S_{\nn}(x,t)$ as
\begin{equation}
  \begin{split}
S_{\nn}(x,t)= -\ln{\epsilon}&-\frac{1}{2}\left< (\phi_{\nn}(x,t)\!-\!\phi_{\nn}(0,0))^2\right>\\
&+\!\frac{1}{2}\left[\phi_{\nn}(x,t),\phi_{\nn}(0,0)\right].
\end{split}
\label{25}
\end{equation}
The thermal average and the commutator appearing in this expression can be
simplified via a mode expansion, by expressing
$\phi_{\nn}(x,t)$ in terms of boson creation and annihilation
operators using Eq.~(\ref{eq:defphiboson}). The boson
correlation functions are independent of $\nn$, and so we drop the edge
index.
We find
\begin{align}
S(x,t)=-\ln{\epsilon} - \int_0^{\infty} \!& \frac{dq}{q}e^{-\epsilon q}
\big(
\coth{\left(\beta\hbar\omega(q)/ 2 \right)}\nonumber\\
&\times\left[1-\cos{(qx-\omega(q) t)}\right] \nonumber\\ 
&-i\sin{(qx -\omega(q) t)}
\big) .
\label{26}
\end{align}
%where the taking the thermodynamic limit was taken ($\beta = 1/k_{\rm B} T$).
%\begin{equation}
%S(x,t)=-\log{\epsilon}-
%\!\!\int_0^{\infty}\!\!\frac{dq}{q}e^{-\epsilon q}
%\!\times\!\!\bigg\{ \!\!\coth{\!\left(\,\beta\hbar\omega_q/\,2\,\right)}\!
%\left[1\!-\!\cos{(qx\!-\omega_q t)}\right]
%+i\sin{(qx\!-\omega_q t)}
%\!\!\bigg\}.
%\label{eq:fullS}
%\end{equation}
%We note that the function $S(x,t)$ and consequently $g_{\nn}(r,t)$ are
%independent of the edge index $\nn$ and therefore we will drop the index
%from now on. It is useful to note that
%\begin{equation}
%G(-x,-t)=G(x,t)^{\ast}\label{eq:Gstar}.
%\end{equation}
%and also to define a frequency-dependent correlator,
%\begin{equation}
%\tilde{g}(x,\Omega)=\int\! \! \! dt \: e^{i\Omega t}g(x,t).
%\end{equation}

Combining these ingredients, we have
\begin{equation}
\langle\psi_{\nn}^{\dagger}(x,t)\psi
\phdagg_{\nn}(0,0)\rangle = \frac{1}{2\pi}e^{i\mu_m(t-x/v)/\hbar} 
e^{S(x,t)}\,.
\label{G}
\end{equation}
By a similar calculation, we find
\begin{equation}
\langle\psi\phdagg_{\nn}(x,t)\psi_{\nn}^{\dagger}(0,0)\rangle = 
\frac{1}{2\pi}e^{-i\mu_m(t-x/v)/\hbar} 
e^{S(x,t)}\,.
\label{G*}
\end{equation}

\section{Conductance and Noise Power}\label{sec:cond}
In this section we express the differential conductance
and noise power for the interferometer
in terms of the correlation function evaluated in Sec.~\ref{ssec:model:G}.
%The voltage term can be removed trough a gauge
%transformation on the fermionic field operators under which,
%\begin{equation}
%\psi^{\dagger}_{n}(x) \rightarrow e^{ i \theta_{n}(x)}
%\psi^{\dagger}_{n}(x),
%\end{equation}
%where
%\begin{equation} 
%\theta_{n}(x)= \frac{\hbar v}  x \mu_{\nn}. 
%\end{equation}
% The phase shift acquired by the means of this transformation generates a
%  phase factor for the tunneling amplitudes
%\begin{eqnarray}}
%\t_{a} \rightarrow t_{a} e^{ - i\left( \theta_1 x_{a1}- \theta_2
%x_{a2} \right) \nonumber \\
%\t_{b} \rightarrow t_{a} e^{ - i\left( \theta_1 x_{b1}- \theta_2
%x_{b2} \right)
%\end{eqnarray}}
%\subsubsection{Electron Tunneling}\label{ssec:cond:kubo}
As a first step, we require the operator
${\hat I}$ representing current from edge 1 to edge 2. 
It can be 
obtained from the
time evolution of the total charge operator, as
\begin{eqnarray}
\label{10}
{\hat I}&=& -e \frac{d}{dt}{\hat N_1}=-\frac{i e}{\hbar}[\Ham,\hat
N_1]=\frac{i e}{\hbar}[\Ham,\hat N_2]  \\
&=&\frac{e}{\hbar} \left\{ 
i t_a \psi^{\dagger}_1(0) \psi\phdagg_2(0) +
i t_b \psi^{\dagger}_1(d_{1}) \psi\phdagg_2(d_{2})+ H.c. \right\}
\, \nonumber . 
%=\frac{e}{\hbar} \left\{ 
%i t_a \frac{e^{i(\mu_1-\mu_2) t/\hbar}}{2 \pi \epsilon} e^{i 2 \pi
%({\hat N_1} x_{a1}- {\hat N_2} x_{a2})/L} \exp[i(\phi_1(x_{a1})-\phi_2(x_{a2}))] \right. & & \nonumber\\ 
%\left. + 
%i t_b \frac{e^{i(\mu_1-\mu_2) t/\hbar}}{2 \pi \epsilon} e^{i 2 \pi
%({\hat N_1} x_{b1}- {\hat N_2} x_{b2})/L} \exp[i(\phi_1(x_{b1})-\phi_2(x_{b2}))]+
%h.c. \right\} \,,
%\frac{ e}{\hbar} \left\{ 
%i t_a\exp(-ieV t/\hbar)\exp[i(\phi_1(x_{a1})-\phi_2(x_{a2}))] \right. & & \nonumber\\ 
%\left. + i t_b\exp(-ieVt/\hbar)\exp[i(\phi_1(x_{b1})-\phi_2(x_{b2}))]+ h.c. \right\} \,
\end{eqnarray}
%where $\Ham=\Ham_{0}+\Ham_{\textrm{int}}+ \Ham_{\rm tun}$.

\subsection{Conductance} 
The expectation value for
the current at time $t$ is 
\begin{equation}
\label{11}
I(t) =\langle {\cal U}(-\infty,t){\hat I}(t) {\cal U}(t,-\infty) \rangle 
\end{equation}
where ${\cal U}(t_2,t_1)$ is the operator for time evolution from time $t_1$
to $t_2$. In the interaction representation that we are using, it
has the form
${\cal U}(t_2,t_1)={\rm T_t}\exp \left(-
  \frac{i}{\hbar}\int^{t_2}_{t_1}\Ham_{\rm tun}(t)  dt
\: \right)$, where $\rm T_t$ denotes 
time-ordering.

%The voltage
%bias $V$ is set by a chemical potential imbalance $\delta \mu=eV/2$
%\begin{equation}
%\Ham_{eV}=  -  \delta \mu \left({\hat N_1} -{\hat
%N_2}  \right).
%\label{27} 
%\end{equation}
%It is convenient to gauge transform the voltage bias into a local but
%time-dependent hopping amplitude by means of a phase transformation on
%the fermionic operators,
%\begin{equation}
%\psi^{\dagger}_{1}(x) \rightarrow e^{ -i \delta \mu t {\hat N}_{1}/\hbar}
%\psi^{\dagger}_{1}(x), 
%\psi^{\dagger}_{2}(x) \rightarrow e^{ i \delta \mu t {\hat N}_{2}/\hbar}
%\psi^{\dagger}_{2}(x),
%\label{28}
%\end{equation}
%where
%\begin{equation} 
%\theta_{1}(t)= - \int_{-\infty}^t\! \! \! dt \: \delta \mu {\hat N}_{1}  /\hbar . 
%\label{29}
%\end{equation}
%The phase shift acquired by the means of this transformation generates
%a phase factor for the tunneling amplitudes
%\begin{equation}
%\rt \rightarrow \rt e^{ - i eV t/hbar }. 
%\end{equation}

To lowest order in the tunneling we have
\begin{equation}
I(V) = -\frac{i}{\hbar} \int_{-\infty}^{0}\! \! \! dt \: \langle \left[ \hat
I(0), \Ham_{\rm tun}(t) \right] \rangle.
\label{30}
\end{equation}
Using the correlation functions of
Eqns.~(\ref{G}) and (\ref{G*}), 
%\begin{equation}
%G_{\nn}(x,t)= e^{i \mu_{\nn} ( t- x/\vO)/\hbar }g(x,t)
%\end{equation} 
%For the purpose of
%this calculation, we will consider that the edge $2$ is connected to
%the ground, i.e. $\mu_1 \rightarrow \mu +eV, \mu_2 \rightarrow \mu$. 
the average current is 
\begin{widetext}
\begin{eqnarray}
\label{12}
I(V) = \frac{-2 e}{\hbar^2}
\int_{-\infty}^{+\infty}dt \:&\Big[ & \left(|t_a|^2+|t_b|^2 \right)
e^{-i(\mu_1 - \mu_2) t/\hbar} (i) \Im  \left[   g(0,t)^2 \right]   \nonumber \\ 
%& &\frac{-2 e}{\hbar^2} \int_{-\infty}^{+\infty}dt \: 
&&+\left\{
t_a t_b^{\ast} e^{i(\mu_1 d_1 - \mu_2 d_2)/\hbar v}
 e^{-i e (\mu_1-\mu_2)t /\hbar}  (i) \Im \left[  g(d_1,t)
g(d_2,t)  \right]  
+ c.c. \right\} \Big] \,. 
\end{eqnarray}

The voltage applied to the interferometer is related to the chemical
potentials of the edges by $V=(\mu_1-\mu_2)/e$.
The differential conductance, $\sigma(V)\equiv d I(V)/dV$,
has a contribution $\sigma_0(V)$ which is independent of the
enclosed flux $\Phi$, and another, $\sigma_\Phi(V)$,
which varies with $\Phi$. Choosing $\mu_2$ independent of $V$, and
introducing $G_0 = e^2/h$, the quantum unit of conductance, we find
at leading order in tunneling
\begin{eqnarray}
\label{12a}
%\sigma(V)&=&\sigma_0(V)+\sigma_\Phi(V), \nonumber \\
\sigma_{0}(V)&=&  -4 \pi \sig \frac{|t_a|^2+|t_b|^2}{ \hbar^2}
\int_{-\infty}^{+\infty}\! \! \! dt \:  e^{-ie V t/\hbar} t \Im
\left[   g(0,t)^2  \right], \nonumber \\
\sigma_{\Phi}(V)&=& -4 \pi \sig  \left\{ \frac{t_a t_b^{\ast}}{\hbar^2}e^{i\mu_2(d_1 - d_2)/\hbar v}
\int_{-\infty}^{+\infty}\! \! \! dt \: e^{-ieV t/\hbar} t \Im \left[ 
 g(d_1,t+d_1/v ) g(d_2,t+d_1/v)  \right]
+ c.c. \right\}\,.
\label{differential-conductance}
\end{eqnarray}
\end{widetext}
This is one of our central results. 
As in a non-interacting system, the differential conductance 
has zeroth and first harmonics in the flux ratio $\Phi/\Phi_0$,
but with interactions their amplitudes acquire a bias dependence,
which enters via the time dependence of the correlation function 
$g(x,t)$. We postpone a discussion of both this bias dependence and of temperature
dependence to Sec.~\ref{sec:visibility}, 
after our analysis in Sec.~\ref{sec:correlation-fn}
of the correlation function. From the structure of the calculation, it
is clear that at higher order in tunneling amplitudes, higher
harmonics appear in the flux-dependence of the differential
conductance as a consequence of interactions

\subsection{Noise power}\label{sec:noise} 

Now we turn to a calculation of the noise power $P(\omega,V)$ at
frequency $\omega$ and voltage $V$, defined in terms of the current by
\begin{equation}
P(\omega,V )= \int_{-\infty}^{\infty} \! \! dt \cos({\omega t}) \left(
  \langle  \hat{I}(t)  \hat{I}(0) \rangle - 
\langle \hat{I}(0) \rangle ^2 \right) \,.
\end{equation}
Note that this includes contributions from both shot noise and Nyquist noise.
At lowest order in the tunneling amplitude we need keep only the
first term. We find
%\begin{equation}
%S(\omega )= \int_{-\infty}^{\infty} \! \! dt e^{i \omega t}  
%\frac{1}{2} \langle \{ I(t) , I(0) \} \rangle 
%\end{equation}
\begin{widetext}
%\begin{eqnarray}
%S(\omega) =\frac{2 e^2}{\hbar^2}
%\int_{-\infty}^{+\infty}dt \cos(\omega t)\, &\Big[&(|t_a|^2+|t_b|^2)
%e^{-ie V t /\hbar} \Re  \left[   g(0,t)^2 \right]   \nonumber \\ 
%%& &+2 e^2 \int_{-\infty}^{+\infty}dt \: 
%&& + \left\{
%t_a t_b^{\ast} e^{i\mu_2(d_1 - d_2)/\hbar v}
%e^{-ie Vt /\hbar}   \Re \left[  g(d_1,t+d_1/v)
%g(d_2,t+d_1/v)  \right]  
%+ h.c. \right\} \Big]\,.
%\label{noise}
%\end{eqnarray}
\begin{eqnarray}
P(\omega,V) =\frac{2 e^2}{\hbar^2}
\int_{-\infty}^{+\infty}dt \cos(\omega t)\, &\Big[&(|t_a|^2+|t_b|^2)
e^{-ie V t /\hbar} \Re  \left[   g(0,t)^2 \right]   \nonumber \\ 
%& &+2 e^2 \int_{-\infty}^{+\infty}dt \: 
&& + \left\{
t_a t_b^{\ast} e^{i(\mu_1 d_1 - \mu_2 d_2)/\hbar v}
e^{-ie Vt /\hbar}   \Re \left[  g(d_1,t)
g(d_2,t)  \right]  
+ c.c. \right\} \Big]\,.
\label{noise}
\end{eqnarray}
\end{widetext}
Hence, at leading order in tunneling, the flux dependence of the noise power
has only zeroth and first harmonics. At higher order, we
expect a second harmonic, as in the non-interacting system, and higher
harmonics, as a consequence of interactions.  

At leading order in tunneling, there is a version of the
fluctuation-dissipation theorem that applies even at finite bias, and
that allows one to relate noise power to current.\cite{scalapino}
It is derived as follows. First, 
%note that
%\begin{equation}
%P(\omega,V)= \frac12 \left( P(0,V+\hbar \omega/e)+ P(0,V - \hbar \omega/e) \right)\,.
%\end{equation}
%Next, 
introduce  the operator $\A= t_a \psi\dagg_1(0) \psi\phdagg_2(0)+t_b \psi\dagg_1(d_1)
\psi\phdagg_2(d_2)$: we have ${\cal H}_{\rm tun} = \A + \A^{\dagger}$
and $\hat{I} = (ie/\hbar)(\A - \A^{\dagger})$.
Then at leading order in tunneling
\begin{equation}
I(V)= -\frac{e}{\hbar^2} \int_{-\infty}^{\infty} dt \langle \left[ \A^{\dagger}(t), \A(0) \right] \rangle\,.
\end{equation}
In a similar fashion, the noise power at leading order is given by
\begin{equation}
P(\omega, V)= \frac{e^2}{\hbar^2} \int_{-\infty}^{\infty} dt \cos(\omega t) \langle \left\{ \A^{\dagger}(t), \A(0) \right\} \rangle.
\end{equation}

At this point it is useful to introduce a spectral decomposition of
$\A$ in terms of the eigenstates of the Hamiltonian
in absence of tunneling. Since the Hamiltonian in the absence of tunneling
conserves the number of particles on each edge, one can use the basis
$|\n \rangle= | E_{\n},N_1,N_2 \rangle$ for this spectral decomposition
and define
%\begin{equation}
%\A(\omega) = 2 \pi \sum_{\n,\m} \left(\rho_\n+\rho_\m \right) |\langle \m |\A |\n \rangle |^2 \delta \left( \hbar \omega+ E_\n - E_\m \right),
%\end{equation}
\begin{eqnarray}
A(\omega) =& 2 \pi Z^{-1}\sum_{\n,\m} \left(e^{-\beta E_\n}+e^{-\beta
    E_\m} \right) |\langle \m |\A |\n \rangle |^2  \nonumber\\&\times
\delta \left( \hbar \omega+ E_\n - E_\m \right),
\end{eqnarray}
where $Z= \sum_{\n} e^{- \beta E_\n}$. By rewriting the current and
noise power in terms of the basis $| \n \rangle $, one can relate
these two quantities to the spectral function:
\begin{eqnarray}
I(V)&=& \frac{e}{\hbar} \tanh \left[ \beta e V /2 \right] A(eV/\hbar)\, , \\
P(\omega,V) &=&\frac{e^2}{ 2 \hbar} \left[ A(eV/\hbar +\omega) +A(eV/\hbar- \omega) \right]\,. 
\end{eqnarray}
%where $eV=\mu_1-\mu_2$. 
We can therefore rewrite the power as
\begin{equation}
P(\omega,V)= \frac{e}{2} \sum_{\pm} \coth \left[\beta \left( eV \pm
      \hbar \omega \right)/2 \right] I (V \pm \hbar \omega/e)\,. 
\end{equation}
In particular, at zero temperature and zero frequency, we find
$P(0,V)/eI(V)=1$. As a result, at weak tunneling, noise power provides
no extra information compared to conductance.

\subsection{Voltage dependence}

From Eq.~(\ref{differential-conductance})  
we see that the dependence on bias voltage $V$ of
conductance is given by Fourier transforms with
respect to time $t$, and with $V$ as the transform variable, 
of combinations of the correlation function $g(x,t)$.
A similar statement holds for noise power from Eq.~(\ref{noise}), but with $V\pm \hbar
\omega/e$ as the transform variable.
For both conductance and noise power, there are two types of contribution:
one independent of flux, which involves only $g(0,t)$; and one harmonic
in $\Phi$, which involves $g(d_1,t+d_1/v)$ and $g(d_2,t+d_1/v)$.
For an interferometer with arm lengths much greater than 
interaction range, there are distinct voltage scales associated with 
these two types of contribution. For definiteness, we discuss
short-range interactions with a range $b$, as introduced above. Then a
characteristic scale in the time-dependence of $g(0,t)$ is $b/v$
with a corresponding voltage scale $\hbar v/eb$. This is the relevant
scale for the flux-independent contributions to conductance and noise
power: it marks a crossover between separate regimes of free electron
behaviour at low and high bias. At low bias, the relevant edge
velocity is $v$, while at high bias the velocity acquires a finite
renormalization from interactions, and is $v-u(0)$. 
In the following, we restrict
ourselves to the low-bias regime, so that flux-independent quantities
have negligible variation with $V$. In this regime, we focus on the voltage dependence
of Aharonov-Bohm oscillations in conductance and noise power, which is
determined by the time dependence of $g(d_1,t+d_1/v)$ and $g(d_2,t+d_1/v)$.
As we show below, with  $x \gg b$, there is a much larger 
characteristic time scale for $g(x,t)$ 
than for $g(0,t)$, and so the voltage
scale relevant to the Aharonov-Bohm oscillations is much smaller than
$\hbar v/eb$.

\section{Asymptotics of correlation function}\label{sec:correlation-fn}

In order to obtain the voltage dependence of Aharonov-Bohm
oscillations in conductance and noise power, we need to evaluate,
for large fixed $|x|$ as a function of $t$, the correlation function
$g(x,t)$, by computing the integral on $q$ in Eq.~(\ref{26}).
We will find that interactions generate $x$-dependent length and
time scales $\ell$ and $t_{\varphi}$, which characterise the width and
duration of an electron wavepacket passing the point $x$ after
injection at the origin. In addition, at finite temperature a
dephasing length $\ell_{\varphi}$ arises. 
It is worthwhile to compare dephasing in the interferometer 
with the situation in a conventional mesoscopic
conductor treated using linear response theory. In the
interferometer, dephasing may arise either from finite bias voltage or
from thermal excitations: it is therefore characterised by two distinct
scales,  $t_{\varphi}$ and $\ell_{\varphi}$. By contrast, within
linear response  dephasing is due only to thermal excitations,
and length and time scales for dephasing are simply related.
Finally, it is convenient to
parametrize temperature in terms of a thermal length $L_{\rm T} \equiv
\hbar v \beta$ defined for the non-interacting system.

%\begin{equation}
%g(x,t) \equiv \frac{1}{2\pi} \exp(S(x,t))
%\label{g}
%\end{equation}
%where
%\begin{eqnarray}
%S(x,t) &=& -\ln(\epsilon) + \int_0^{\infty} \frac{dq}{q} e^{-\epsilon
%  q}\left\{\coth(\beta \hbar \omega(q)/2)\right.\nonumber\\
%&\times&\left.[\cos(qx-\omega(q)t)-1] + i
%  \sin(qx - \omega(q)t)\right\}\,,
%\label{S}
%\end{eqnarray}
%and $\omega(q)$ is the dispersion relation for bosonic modes.
At large $|x|$ the dominant contribution to
Eq.~(\ref{26}) is from small $q$, and so we are concerned with the form
of $\omega(q)$ at small $q$. We consider separately the cases of short range
interactions, and of Coulomb interactions. 
For the first, we have
%we have $\omega(q) = q[v + u(q) - u(0)]$, where $u(q)$ is the Fourier
%transform of the electron-electron interaction potential. Introducing
%a length $b$ which characterises a combination of the strength and the
%range
%of interactions, we have the expansion
$$
\omega(q) = v q - (v/b)\cdot (b q)^n \ldots\,,
$$ 
%where $n=3$ arises naturally from Taylor expansion of $u(q)$, while $n=2$
%may appear as a consequence of curvature in the underlying fermion
with $n=3$. We obtain (see Eq.~(\ref{sr-disp})) $\ell
\equiv b(x/b)^{1/n}$, with
\begin{equation}
t_{\varphi}\equiv\ell/v
\label{t-phi-sr}
\end{equation}
and
\begin{equation}
\ell_{\varphi} = L_{\rm T} (L_{\rm T}/b)^{n-1}\,.
\label{ell-phi-sr}
\end{equation}
To understand the physical origin of the dependence of $t_{\varphi}$
on $x$, consider an electron wavepacket of width $\ell$. It consists
of modes with wavevectors in a range from $0$ to ${\cal O}(1/\ell)$. The phase difference
accumulated due to dispersion, between modes at opposite ends of this range, during
propagation over the distance $x$, is $(v/b)\cdot (b/\ell)^n
\cdot (x/v)$, and the value of $\ell$ may be extracted from the
consistency requirement that this phase difference is ${\cal O}(1)$.
%set the characteristic scales for the Green function. 

In the case of Coulomb interactions, we have 
$$
\omega(q) = q[v + u
\ln(1/aq)]\,.
$$ 
We find (see Eq.~(\ref{coulomb-disp})) $\ell = x/\ln(x/a)$, 
\begin{equation}
t_{\varphi} = \ell/[u \ln(\ell/a)]
\label{t-phi-coulomb}
\end{equation}
and
\begin{equation}
\ell_{\varphi} = (u L_{\rm T} /v)\cdot \ln^2(L_{\rm T}/a)\,.
\label{ell-phi-coulomb}
\end{equation}

In addition to dephasing of this kind, due to interactions, 
there may be dephasing of collective bosonic modes 
that arises from curvature of the single-particle dispersion
relation for edge electrons, which was omitted from Eq.~(\ref{eq:H0psi}).
Such curvature generates cubic and higher order terms in the bosonic
Hamiltonian, and hence scattering between collective modes.\cite{haldane}
We estimate the dephasing length $\ell_{\varphi}^{\rm curv}$ arising
from this mechanism as follows.
First, we parameterise curvature in the electron dispersion in terms
of a lengthscale $|D|$, by writing in place of Eq.~(\ref{eq:H0psi})
$$
{\cal H}_0 = \hbar v \sum_{qm} \left(q + \frac{l_{\rm B}^2 q^2}{D} +
\ldots\right) c^\dagger_{qm} c^{\phantom\dagger}_{qm}\,,
$$
where $l_{\rm B}$ is the magnetic length. Following the discussion
of a compressible quantum Hall edge given in Ref.~\onlinecite{Chklovskii},
one expects $|D|$ to be of order the depletion length.
%, which has the value 200nm for typical sample parameters. 
A consequence of this curvature is of course that electrons 
travel at different speeds according to their energy. 
Thermally excited electrons occupy a wavevector range $\Delta q \sim
L_{\rm T}^{-1}$, and hence have a velocity range $\Delta v \sim v
l_{\rm B}^2/(|D| L_{\rm T})$. The dephasing length
is the distance that electrons with wavevectors differing by $\Delta
q$ must propagate to acquire a phase difference ${\cal O}(\pi)$, and so
\begin{equation}
\ell_{\varphi}^{\rm curv} = \frac{v}{\Delta q \Delta v} = |D|\left(\frac{L_{\rm
    T}}{l_{\rm B}}\right)^2\,.
\label{curv}
\end{equation}
Equivalent results have been derived previously
for a non-chiral system, by treating cubic interactions
between bosonic modes perturbatively,\cite{samokhin}
and a discussion including the effects of disorder is
given in Ref~\onlinecite{gornyi}.

Let us summarise the outcome of our discussion
of dephasing lengths important for the correlation function.
Short range interactions ($n=3$ in Eq.~(\ref{ell-phi-sr}))
give
\begin{equation}
\ell_{\varphi} \propto T^{-3}
\end{equation}
while, from Eq.~(\ref{curv}), curvature in the fermion disperison relation leads to
\begin{equation}
\ell_{\varphi}^{\rm curv} \propto T^{-2}\,.
\end{equation}
With short-range interactions, the effects of curvature are therefore dominant in the low-temperature
limit, but $\ell_{\varphi} < \ell_{\varphi}^{\rm curv}$ and
interaction effects dominate above the temperature at which $L_{\rm T} = |D| (b/l_{\rm B})^2$.
By contrast, for Coulomb interactions, from Eq.~(\ref{ell-phi-coulomb}), 
\begin{equation}
\ell_{\varphi} \propto T^{-1} \ln^2(1/T)\,,
\end{equation}
and so
dephasing from interactions dominates over curvature effects
at low temperatures. We discuss numerical estimates for
dephasing lengths in Sec.~\ref{discussion}.

Moving to details of the evaluation of the asymptotic form at large $x$ of the correlation function
for an interacting system with linear electron dispersion, 
there are two separate regimes, according to whether $|x-v t|$ is
small or large compared to $\ell$.

\subsection{Far from the peak: $|x - v t|$ large}

Then it is sufficient to approximate the dispersion relation as
$\omega(q) = v q$, and one obtains the correlation function for
non-interacting particles with speed $v$, which is
\begin{equation}
g(x,t) = \frac{i \pi k_{B} T /\hbar v }{2 \pi \sinh([x - v t]\pi
  k_{B} T/\hbar v)}\,.
\label{free-g}
\end{equation}

\subsection{Near the peak: $|x - v t|$ small}

In this regime it is necessary to take account of the leading
correction at small $q$ to a strictly linear dispersion relation,
which we do separately for short-range and for Coulomb interactions.

\subsubsection{Short-range interactions}

We write
$t=\Delta t + x/v$,  so that at leading order
$$
q x - \omega(q) t = \frac{x}{b} (b q)^n - q v \Delta t \ldots
$$
and rescale variables, using  $\ell$ and $t_{\varphi}$ to define the dimensionless combinations
\begin{eqnarray}
\ell q  &\equiv& Q \,,
\qquad 
\Delta t/t_{\varphi} \equiv \tau\,,
\nonumber \\
\epsilon /\ell&\equiv& \epsilon^{\prime}\,,
\qquad
\beta\hbar /t_{\varphi}\equiv \beta^{\prime}\,.
\nonumber
\end{eqnarray}
Then 
\begin{equation}
\label{sr-disp}
q x - \omega(q) t = (\ell q)^n - \ell q  \Delta t/t_{\varphi} \ldots\,.
\end{equation}
We obtain from Eq.~(\ref{26})
\begin{eqnarray}
S(x,t) &=& -\ln(\ell) - \ln(\epsilon^{\prime})\nonumber\\
&+& \int_0^{\infty} \frac{dQ}{Q} e^{-\epsilon^{\prime} Q}
\left\{\coth(\beta^{\prime} Q/2)\right.\nonumber\\
&\times&\left.[\cos(Q^n - \tau Q)-1] + i
  \sin(Q^n - \tau Q)]\right\}\,.\nonumber
\end{eqnarray}
We can remove the $\epsilon^{\prime}$-dependence by using the result
that, as $\epsilon^{\prime} \rightarrow 0$,
\begin{eqnarray}
- \ln(\epsilon^{\prime}) &+&
\int_0^{\infty} \frac{dQ}{Q} e^{-\epsilon^{\prime} Q} 
\coth( \beta Q / 2)[\cos(Q) - 1]\nonumber\\  
&=& 
\ln\left[\frac{\pi/\beta}{\sinh(\pi/\beta)}\right]
\end{eqnarray}
to write
\begin{eqnarray}
S(x,t) &=& 
-\ln(\ell) +
\ln\left[\frac{\pi/\beta^{\prime}}{\sinh(\pi/\beta^{\prime})}\right] 
\nonumber\\
&+&\int_0^{\infty} \frac{dQ}{Q} 
\big\{\coth(\beta^{\prime} Q/2)\times\nonumber\\
&&[\cos(Q^n - \tau Q)-\cos(Q)] \nonumber\\
&&+ i
  \sin(Q^n - \tau Q)]\big\}\,.
\label{Sfinal}
\end{eqnarray}
From the asymptotic behaviour of Eq.~(\ref{Sfinal}) at large $|\tau|$, one can recover
the form for the correlation function at large $|x-vt|$ given in Eq.~(\ref{free-g}).

The analysis leading to Eq.~(\ref{Sfinal}) makes clear the scales that are important
for $g(x,t)$. 
Considered as a function of $t$ at large, 
fixed $x$,  $g(x,t)$ has a peak near $t=x/v$. The peak has a width in $t$
of order $\ell/v$ and near the peak
$|g(x,t)| \sim \ell^{-1}$.
%At this stage it is useful to introduce a thermal length 
%$L_{\rm  T}=\hbar v \beta$ defined for the noninteracting system.
The interaction energy scale that enters the temperature or voltage dependence of
interferometer fringe visibility is $\hbar  /t_{\varphi}$.
Correspondingly, at a given inverse temperature $\beta$, fringes are visible
only for $\ell \leq \beta \hbar v$, from which we see that the
dephasing length is as given in Eq.~(\ref{ell-phi-sr}).
To find the detailed asymptotic form of the correlation function, it is necessary to
evaluate Eq.~(\ref{Sfinal}) numerically.

\subsubsection{Unscreened Coulomb interactions}

Starting from
$
\omega(q) = q[v_0 + u \ln(1/aq)]\,,
$
we would again like a simplified asymptotic form for
the correlation function 
at large $|x|$ and $|t|$. 
Without screening, the phase velocity is $v +u \ln(1/aq)$
and hence divergent at small $q$.
Focussing on behaviour near the
peak of the correlation function, we
write
$
t = \Delta t + x/[v + u \ln(x/a)]
$
so that
$$
q x - \omega(q) t = \frac{qux\ln(qx)}{v+u\ln(x/a)} - q[v +
u\ln(1/qa)]\Delta t\,.
$$
We assume for simplicity that $u \ln(x/a) \gg v$ (though this is not
essential)
and introduce the characteristic length $\ell = x/\ln(x/a)$
and time $t_{\varphi} = \ell/[u \ln(\ell/a)]$. Then
\begin{equation}
\label{coulomb-disp}
qx - \omega(q)t = \ell q \ln(\ell q) - \ell q \Delta t/t_\varphi \ldots\,.
\end{equation}
Defining dimensionless variables as before,
the asymptotic form of the correlation function is 
obtained from
\begin{eqnarray}
S(x,t) &=& -\ln(\ell) + 
\ln\left[\frac{\pi/\beta^{\prime}}{\sinh(\pi/\beta^{\prime})}\right]
\nonumber\\
&+&
\int_0^{\infty} \frac{dQ}{Q} \big\{ 
\coth(\beta^{\prime} Q/2)\times\nonumber\\
&&[\cos(Q\ln(Q) - Q \tau) -\cos(Q)]\nonumber\\
&&+\left. i \sin(Q\ln(Q) - Q \tau)
\right\}\,.
\label{sfinal-coulomb}
\end{eqnarray}
In this way we see for Coulomb interactions, taking the energy scale for
visibility of interference fringes as $\hbar/t_{\varphi}$, that the dephasing
length is as in Eq.~(\ref{ell-phi-coulomb}).

%\begin{widetext}

\subsection{Numerical evaluation of the asymptotic form for the correlation
  function}

\begin{figure}[ht]
\includegraphics[width=8cm]{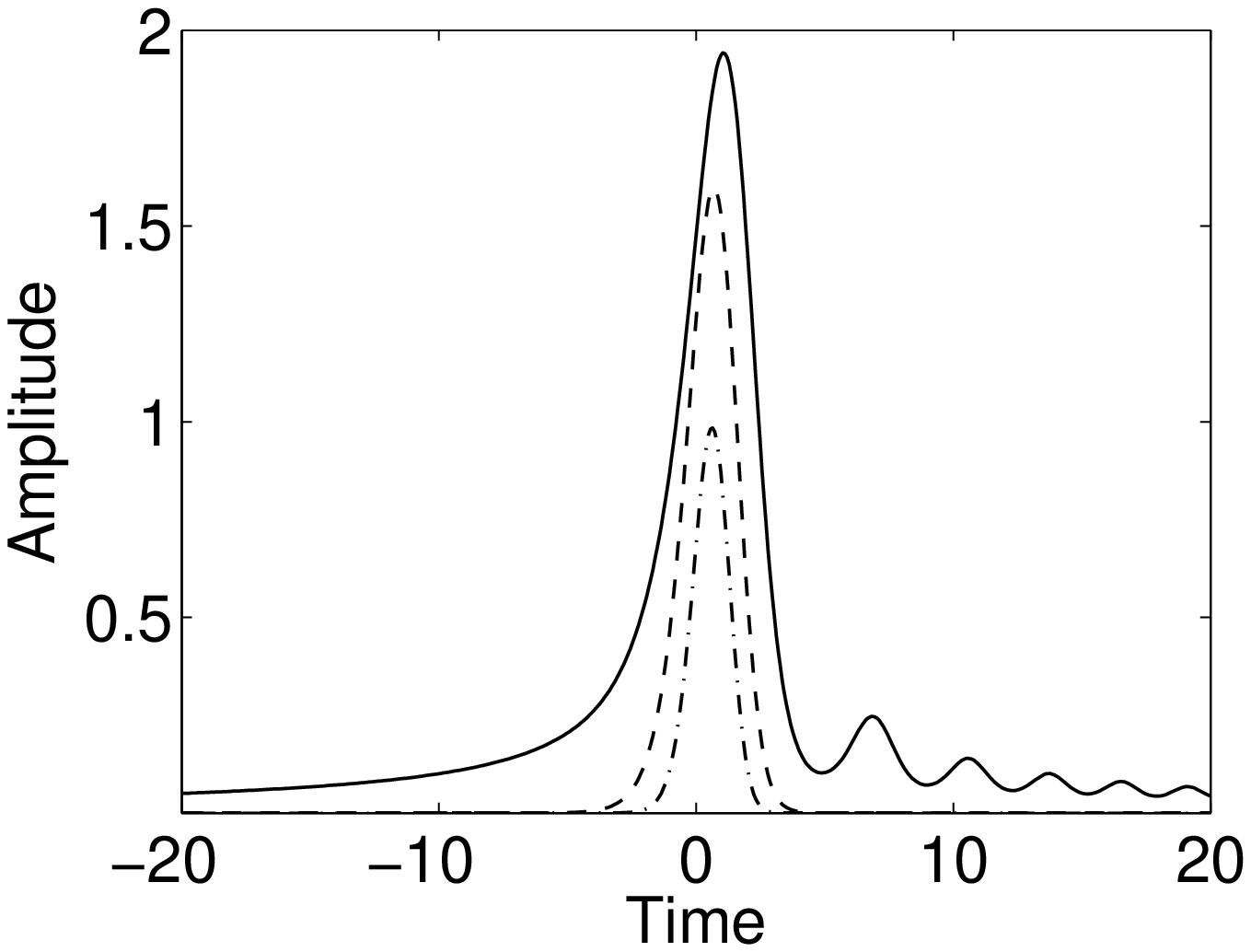}
\includegraphics[width=8cm]{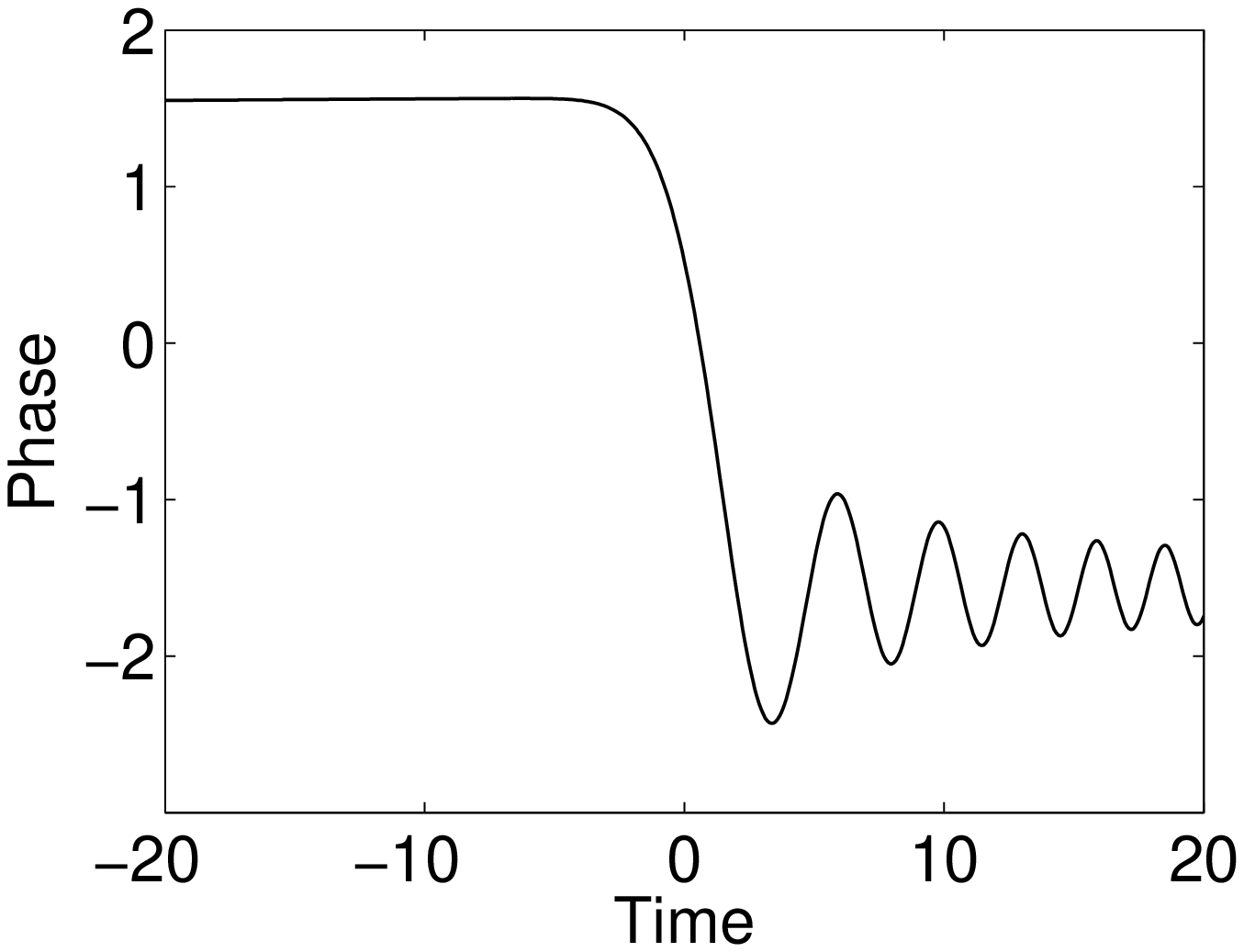}
\caption{Asymptotic behaviour of the correlation function
for short range interactions. Upper panel: the function $\exp(U(\beta^{\prime},\tau))$, proportional to its
modulus, as a function of $\tau$, for $\beta^{\prime} = \infty$ (full line),
$\beta^{\prime}=2$ (dashed line), and $\beta^{\prime}=1$ (dash-dotted line). Lower panel: its phase,
$W(\tau)$.} 
\label{g-short-range}
\end{figure}

\begin{figure}[ht]
\includegraphics[width=8cm]{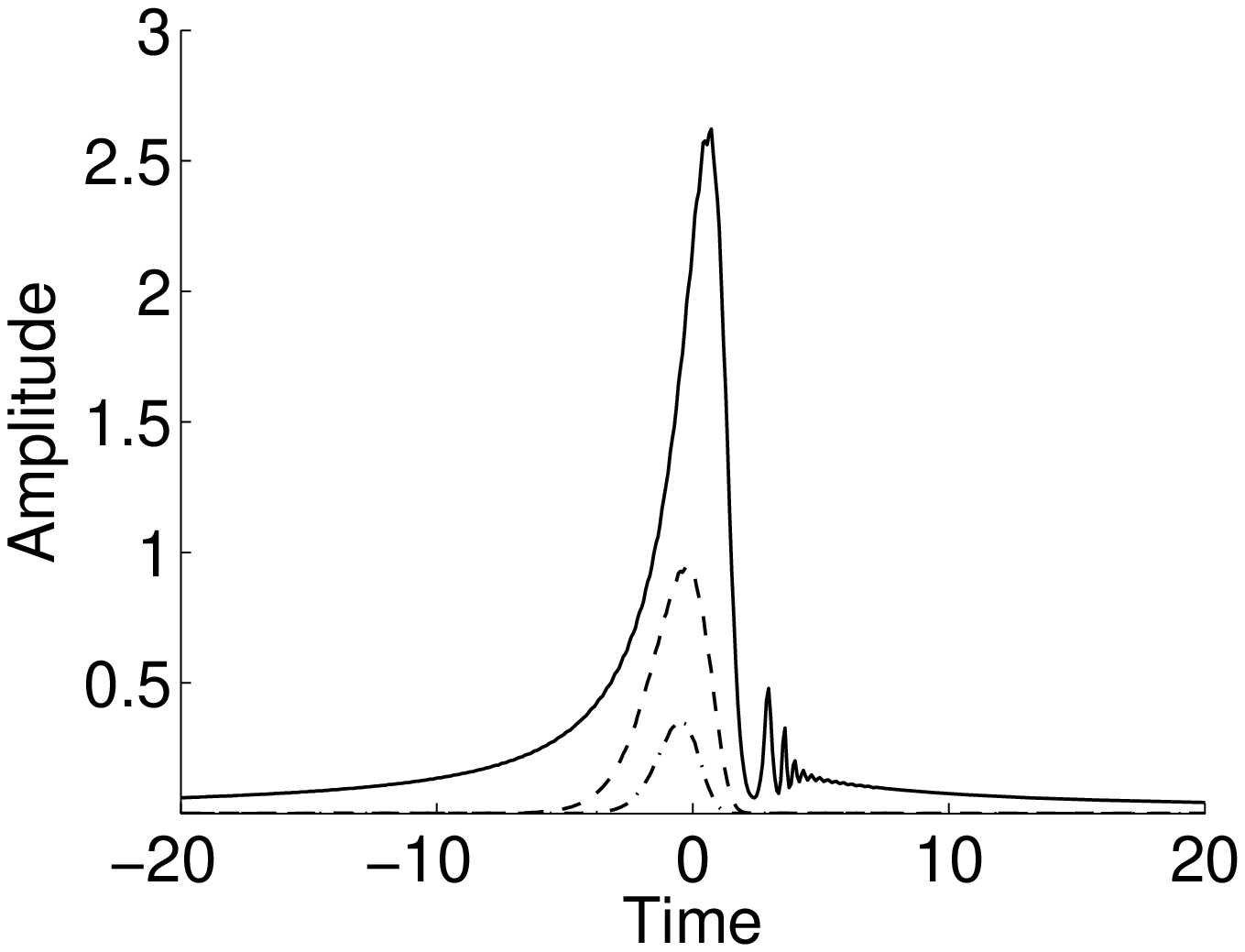}
\includegraphics[width=8cm]{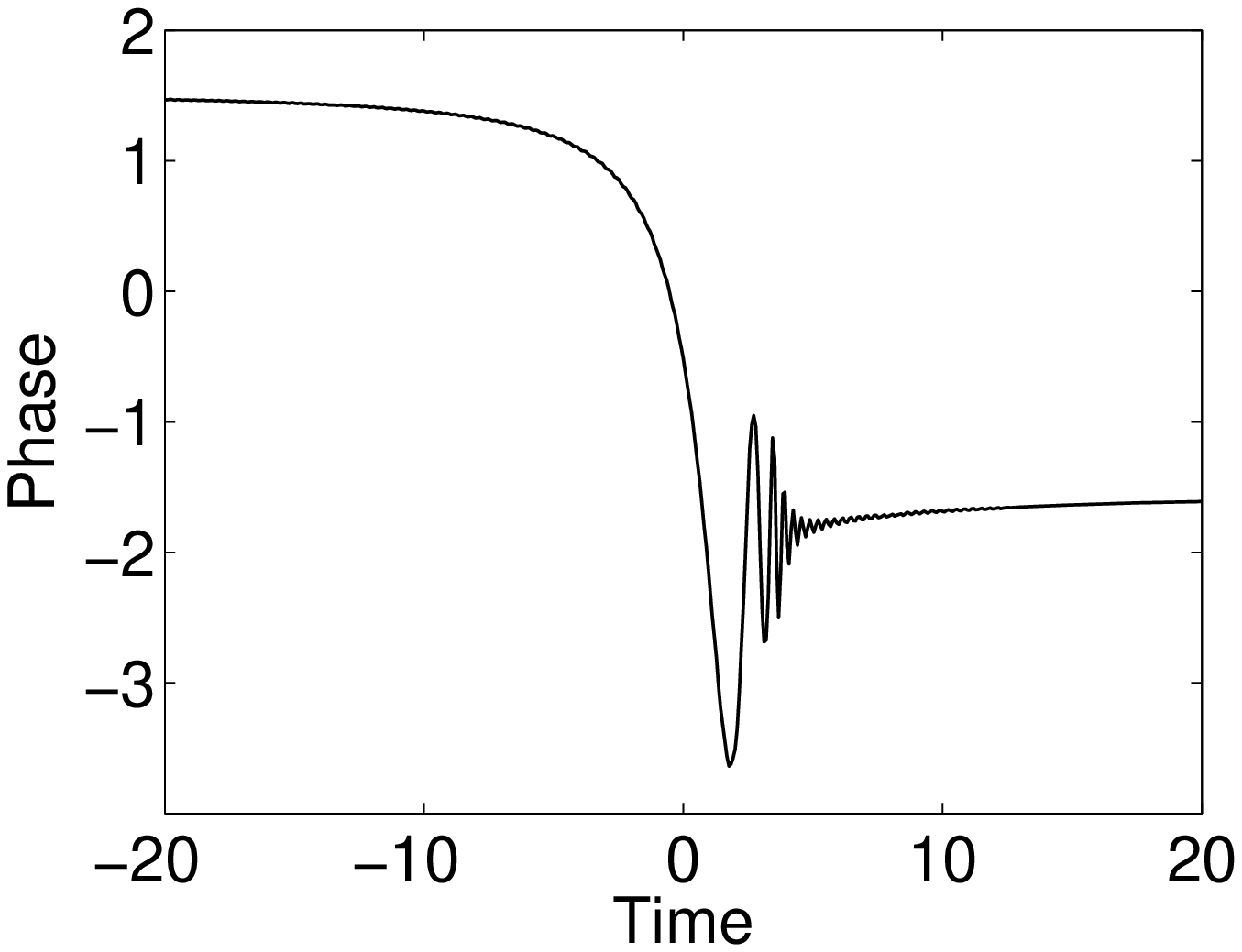}
\caption{Asymptotic behaviour of the correlation function
for Coulomb interactions. Upper panel: the function $\exp(U(\beta^{\prime},\tau))$, proportional to its
modulus, as a function of $\tau$, for $\beta^{\prime} = \infty$ (full line),
$\beta^{\prime}=2$ (dashed line), and $\beta^{\prime}=1$ (dash-dotted line). Lower panel: its phase,
$W(\tau)$.} 
\label{g-coulomb}
\end{figure}

For both short-range interactions and Coulomb interactions it is convenient
to write
$$
S(x,t) = -\ln(\ell)  
%\ln\left[\frac{\pi/\beta^{\prime}}{\sinh(\pi/\beta^{\prime})}\right]
+ U(\beta^{\prime},\tau) + i W(\tau)
$$
where $U(\beta^{\prime},\tau)$ and $W(\tau)$ are each real, and can be
read off from Eqns.~(\ref{Sfinal}) and (\ref{sfinal-coulomb}).
%
%\begin{eqnarray}
%U(\beta^{\prime},\tau) = 
%\ln\left[\frac{\pi/\beta^{\prime}}{\sinh(\pi/\beta^{\prime})}\right]+
%\left\{ 
%\begin{array}{ll}
%\int_0^{\infty} \frac{dQ}{Q} \coth(\beta^{\prime}Q/2)
%\left[\cos(Q^n - \tau Q)-\cos(Q) \right] &\qquad {\rm short-range}\\
%&\\
%\int_0^{\infty} \frac{dQ}{Q} \coth(\beta^{\prime}Q/2)
%\left[\cos(Q\ln(Q) - \tau Q)-\cos(Q)] 
%\right] &\qquad {\rm Coulomb}
%\end{array}
%\right.\nonumber
%\end{eqnarray}
%and
%\begin{eqnarray}
%W(\tau) = \left\{ 
%\begin{array}{ll}
%\int_0^{\infty} \frac{dQ}{Q} 
%  \sin(Q^n - \tau Q) &\qquad {\rm short-range}\\
%& \\
%\int_0^{\infty} \frac{dQ}{Q} 
%  \sin(Q\ln(Q) - \tau Q) &\qquad {\rm Coulomb}
%\end{array}
%\right.\nonumber
%\end{eqnarray}
%\end{widetext}
The modulus of the correlation function is proportional to $\exp(U(\beta^{\prime},\tau))$,
and its phase is $W(\tau)$. These functions are shown in
Fig.~\ref{g-short-range} for short-range
interactions with $n=3$, and in Fig.~\ref{g-coulomb} for Coulomb interactions.
%Fig.~\ref{mod-g} and
%Fig.~\ref{phase-g}.
%\begin{figure}[ht]
%\includegraphics[width=6cm]{mod-g.eps}
%\caption{The function $\exp(U(0,\tau))$, which is proportional to the
%modulus of the Green function, as a function of $\tau$ for
%short-range interactions with $n=3$} 
%\label{mod-g}
%\end{figure}
%\begin{figure}[ht]
%\includegraphics[width=6cm]{phase-g.eps}
%\caption{The function $V(\tau)$, which is the phase of the Green
%function, as a function of $\tau$, for short-range interactions with $n=3$} 
%\label{phase-g}
%\end{figure}
%
%The corresponding functions for Coulomb interactions are shown in
%Fig.~\ref{mod-g-log} and Fig.~\ref{phase-g-log}.
%\begin{figure}[ht]
%\includegraphics[width=6cm]{mod-g-log.eps}
%\caption{The function $\exp(U(0,\tau))$, which is proportional to the
%modulus of the Green function, as a function of $\tau$ for 
%Coulomb interactions} 
%\label{mod-g-log}
%\end{figure}
%\begin{figure}[ht]
%\includegraphics[width=6cm]{phase-g-log.eps}
%\caption{The function $V(\tau)$, which is the phase of the Green
%function, as a function of $\tau$ for Coulomb interactions} 
%\label{phase-g-log}
%\end{figure}

\section{Visibility of interference fringes}\label{sec:visibility}

Coherence in the interferometer can be characterised by the
interference fringe visibility, defined as
\begin{equation}
{\cal V} = \frac{{\rm max}_\Phi \sigma(V) - {\rm min}_\Phi \sigma(V)}
{{\rm max}_\Phi \sigma(V) + {\rm min}_\Phi \sigma(V)}\,.
\end{equation}
To provide a context for discussion of interaction effects, it is
useful to recall the result of single-particle theory
for the source-drain transmission probability of electrons
with wavevector $k$. In terms of transmission and reflection
amplitudes $\tau_{a,b}$ and $r_{a,b}$ at the point contacts $a$ and $b$, this
is\cite{heiblum1}
\begin{equation}
|\tau_a r_b e^{ikd_2} + r_a \tau_b e^{ikd_1}|^2\,.
\end{equation}
It is characterised by the energy scale $E_c \equiv \hbar v/|d_1 -
d_2|$, which is divergent for an interferometer with equal arm
lengths. In a non-interacting system, the amplitude of Aharonov-Bohm 
oscillations in the differential conductance of the interferometer
is independent of bias $V$, and has a phase $eV/E_c$. Oscillations are
suppressed at high temperature 
by thermal smearing with a characteristic temperature scale set by
$E_c$, and the visibility is
\begin{equation}
{\cal V}=\frac{2|\tau_a \tau_b r_a r_b|}{|\tau_a|^2|r_b|^2+|\tau_b|^2|r_a|^2}
\times\frac{\pi k_{\rm B} T/E_c}{\sinh(\pi k_{\rm B} T/ E_c)}\,. 
\end{equation}

Turning to interaction effects, we find in general at leading order
in tunneling that
\begin{equation}
{\cal V} = \frac{2|t_a t_b|}{|t_a|^2 + |t_b|^2}\times {\cal V}_{\rm rel}(V^{\prime},\beta^{\prime})\,,
\end{equation}
where $V^{\prime}$ and $\beta^{\prime}$ are scaled voltage
and inverse temperature, respectively.
In the limit of vanishing temperature and voltage, nonlinearity in the
dispersion relation for bosonic excitations is not probed and so
${\cal V}_{\rm rel}(0,\infty)=1$; away from this limit we compute its
form numerically. Because the asymptotic form of the correlation
function depends on interactions only via the functional form of
$u(q)$ at small $q$ and the values of $\ell$ and
$\tau_{\varphi}$, numerical calculations yield universal results
as a function of scaled voltage and temperature. These
results apply provided the interferometer arms are much
longer than the interaction length $b$. It is also possible
to evaluate our expressions numerically
for shorter arm lengths, without using the discussion of 
Sec.~\ref{sec:correlation-fn}: our exploratory
studies yielded results similar to the ones for the asymptotic regime
of large arm length, presented below.

We discuss separately the two cases of interferometers with
equal or unequal arm lengths. In each case, as noted at the end of
Sec.~\ref{sec:cond}, for an interferometer with arm lengths large
compared to the interaction range, $\sigma_0(V)$ is almost
constant over the bias range of interest, so that the visibility
of interference fringes in conductance is determined directly by
$\sigma_\Phi(V)$.

\subsection{Interferometer with arms of equal length}

\subsubsection{Zero temperature}

From Eq.~\ref{differential-conductance}, the
oscillatory part of the conductance can be written as
$$
\sigma_{\Phi}(V) = A(\sigma_1(V) + i \sigma_2(V) + c.c.
$$
where $A = -2G_0 \ell t_a t^*_b /(v\hbar)^2$ and
\begin{eqnarray}
&&\sigma_1(V^{\prime}) + i \sigma_2(V^{\prime})\nonumber\\
&=&\int_{-\infty}^{\infty} d\tau \exp(-iV^{\prime} \tau)
\tau \exp(2U(\beta^{\prime},\tau)) \sin(2W(\tau))\,,\nonumber
\end{eqnarray}
with the scaled voltage $V^{\prime}{=}eVt_{\varphi}/\hbar$. 
The amplitude of oscillations has modulus $\sigma(V) = \sqrt{\sigma_1^2(V) +\sigma_2^2(V)}$. 
%We see that we are concerned with the Fourier transform of 
%$\tau \exp(2U(\beta^{\prime},\tau)) \sin(2V(\tau))$: graphs of this are given for
%short-range interactions in
%Fig.~\ref{FTarg} and for Coulomb interactions in Fig.~\ref{FTarg-log}
%\begin{figure}[ht]
%\includegraphics[width=6cm]{FTarg.eps}
%\caption{The function $\tau \exp(2U(0,\tau)) \sin(2W(\tau))$, calculated
%for short-range interactions with $n=3$: its
%Fourier transform gives the voltage dependence of the amplitude of
%Aharonov-Bohm oscillations in  the conductance, for an interferometer 
%with arms of equal length at zero temperature} 
%\label{FTarg}
%\end{figure}
%\begin{figure}[ht]
%\includegraphics[width=6cm]{FTarg-log.eps}
%\caption{The function $\tau \exp(2U(0,\tau)) \sin(2V(\tau))$, calculated
%for Coulomb interactions: its
%Fourier transform gives the voltage dependence of the amplitude of
%Aharonov-Bohm oscillations in  the conductance, for an interferometer
%with arms of equal length at zero temperature} 
%\label{FTarg-log}
%\end{figure}

\begin{figure}[ht]
\includegraphics[width=8cm]{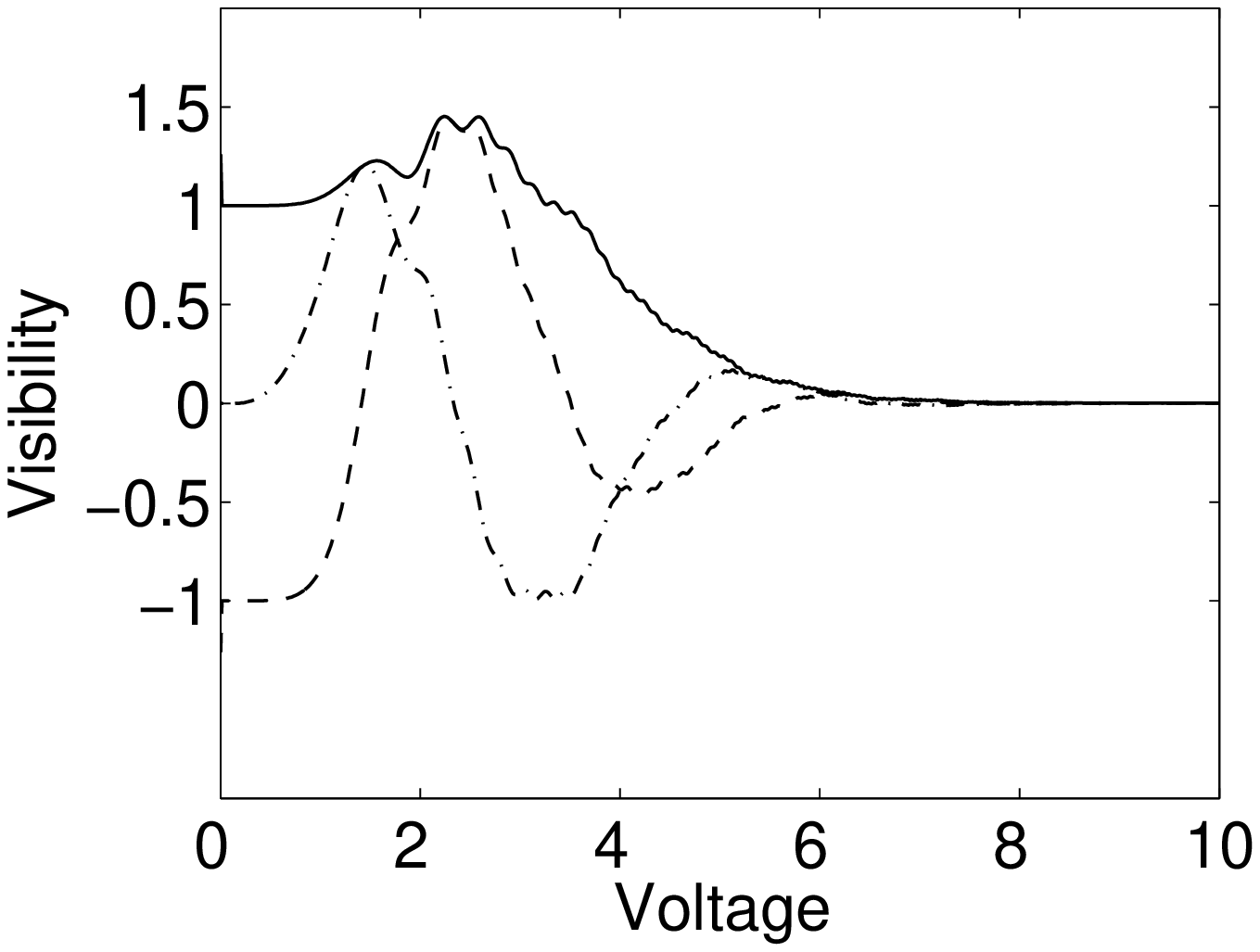}
\includegraphics[width=8cm]{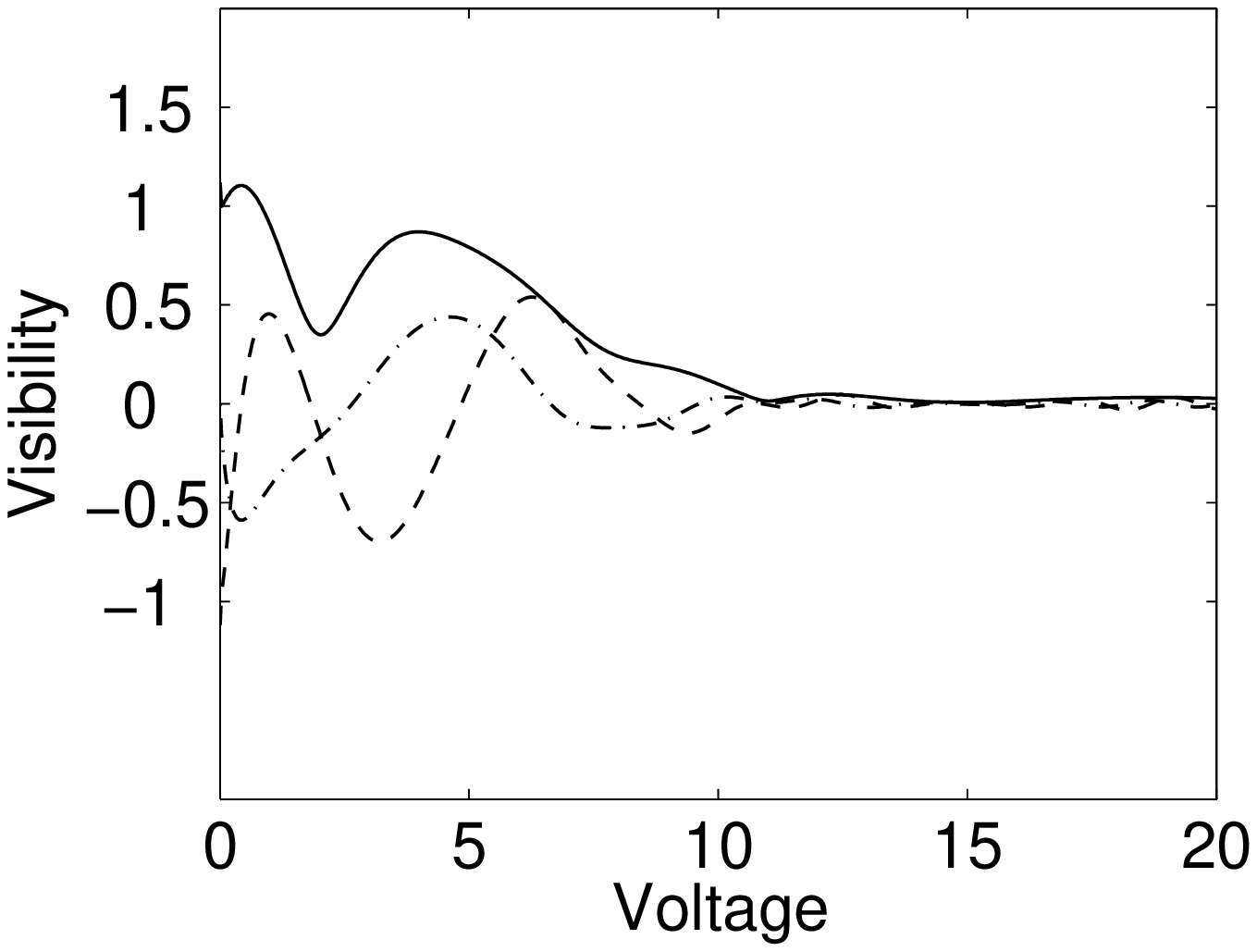}
\caption{The voltage dependence of visibility
for equal length arms at zero temperature.
Upper panel: with short-range interactions; lower panel:
with Coulomb interactions.
Full lines: ${\cal V}_{\rm rel}(V^{\prime},\beta^{\prime})$; dashed lines and dot-dashed
lines: contributions
from $\sigma_1(V)$ and 
$\sigma_2(V)$, respectively.} 
\label{zeroTconductance}
\end{figure}

The resulting visibility ${\cal V}_{\rm rel}(V^{\prime},\beta^{\prime})$, and the
separate contributions derived from
$\sigma_1(V)$ are $\sigma_2(V)$,
are shown in Fig.~\ref{zeroTconductance}. A striking feature is that
for intermediate $V$, the visibility at low temperature exceeds
unity: this means that in this bias range, for some values of $\Phi$,
the differential conductance is negative. 
%\begin{figure}[ht]
%\includegraphics[width=6cm]{sigma-vs-v.eps}
%\caption{The voltage dependence of the visibility of
%Aharonov-Bohm oscillations in  the conductance for equal length arms at zero temperature
%with short-range interactions and $n=3$.
%Full line: $\sigma(V)$; dashed line: $\sigma_1(V)$; 
%dot-dashed line:  $\sigma_2(V)$.} 
%\label{zeroTconductance}
%\end{figure}
%The same quantities calculated for a system with Coulomb interactions
%are shown in Fig.~\ref{conductance-vs-v3}
%\begin{figure}[ht]
%\includegraphics[width=7cm]{conductance-vs-v3.eps}
%\caption{The voltage dependence of the amplitude of
%Aharonov-Bohm oscillations in  the conductance, 
%for equal length arms at zero temperature
%with Coulomb interactions.
%Full line: $\sigma(V)$; dashed line: $\sigma_1(V)$; 
%dot-dashed line:  $\sigma_2(V)$.} 
%\label{conductance-vs-v3}
%\end{figure}

\subsubsection{Non-zero temperature}

\begin{figure}[ht]
\includegraphics[width=8cm]{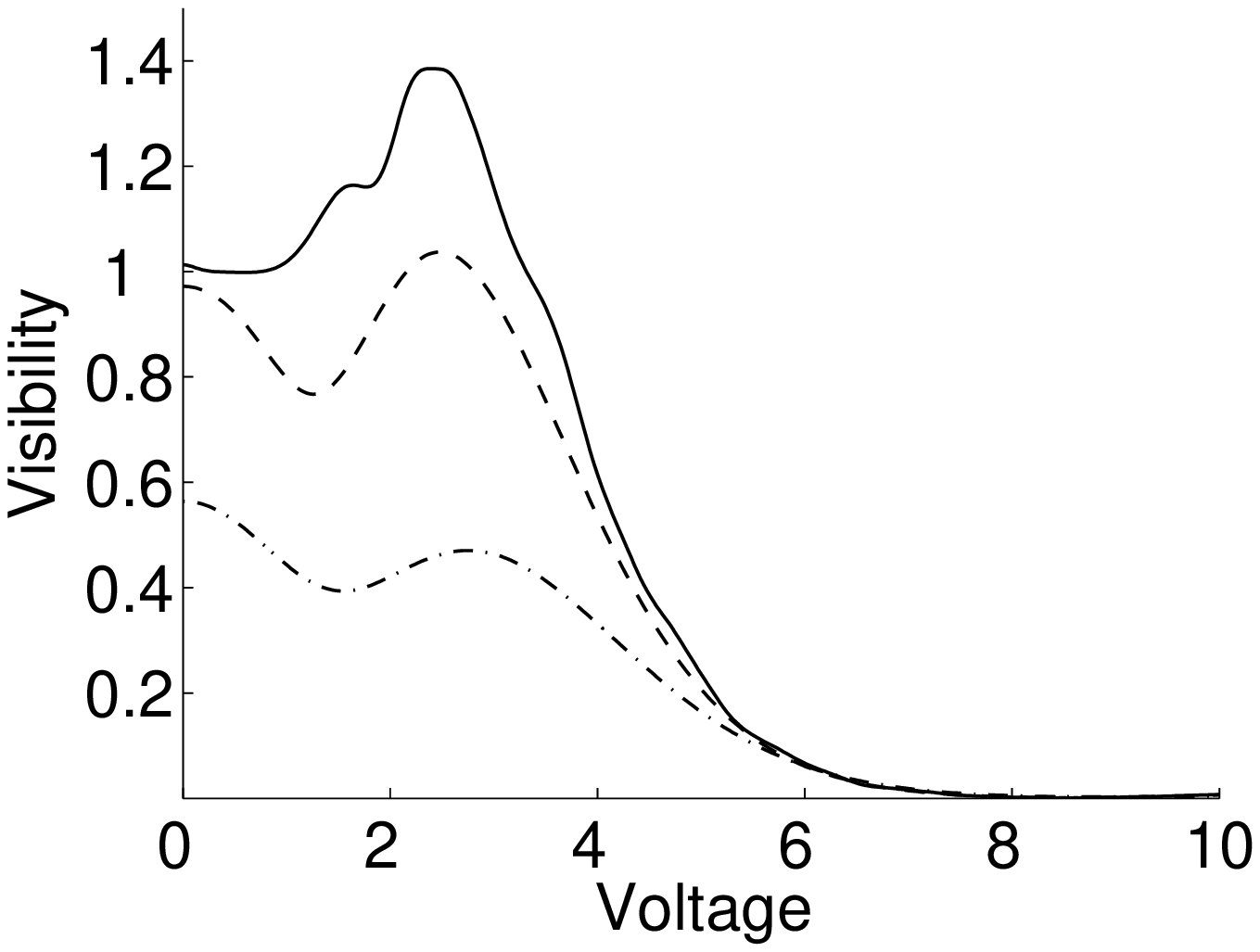}
\includegraphics[width=8cm]{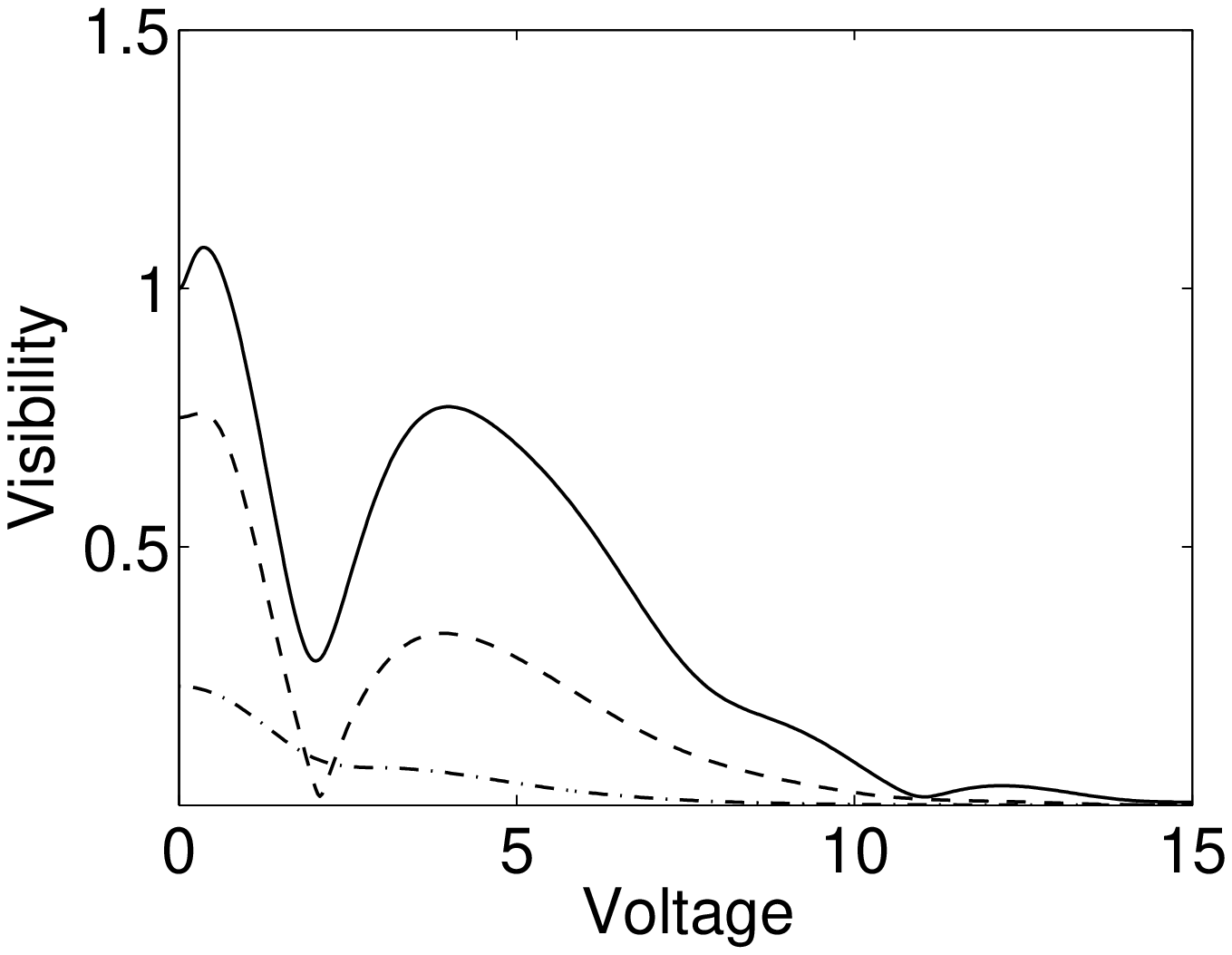}
\caption{The dependence of ${\cal V}_{\rm
    rel}(V^{\prime},\beta^{\prime})$ on $V^{\prime}$,
for equal arm lengths,
at $\beta^{\prime}=20$ (full line),
$\beta^{\prime}=5$ (dashed line), and $\beta^{\prime} = 2$ (dot-dashed
line). Upper panel: short-range interactions; lower panel: Coulomb interactions.} 
\label{b=2-conductance}
\end{figure}

Temperature enters these calculations via the
$\beta^{\prime}$-dependence
of $U(\beta^{\prime},\tau)$. The effect of this
on the modulus of the Green function is shown 
in Fig.~\ref{g-short-range} and \ref{g-coulomb}.
%\begin{figure}[ht]
%\includegraphics[width=6cm]{Tmod-g.eps}
%\caption{The function $\exp(U(\beta^{\prime},\tau)$, which is
%  proportional to the modulus of the Green function, vs.
%$\tau$ for short-range interactions with $n=3$ at $\beta^{\prime}=20,2,1$ and $0.5$.} 
%\label{Tmod-g}
%\end{figure}
The resulting 
%forms of $\sigma_1(V)$, $\sigma(2(V)$ and 
visibility is shown  %$\beta^{\prime} = 2$
in Fig.~\ref{b=2-conductance}. 
%and at $\beta^{\prime} = 1$
%in Fig.~\ref{b=1-conductance}.
%\begin{figure}[ht]
%\includegraphics[width=6cm]{b=1-sigma-vs-v.eps}
%\caption{The voltage dependence of the amplitude of
%Aharonov-Bohm oscillations in  the conductance, for equal arm lengths
%and short-range interactions at $\beta^{\prime}=1$.
%Full line: $\sigma(V)$; dashed line: $\sigma_1(V)$; 
%dot-dashed line:  $\sigma_2(V)$.} 
%\label{b=1-conductance}
%\end{figure}
%The temperature dependence of the conductance with Coulomb
%interactions is shown in Fig.~\ref{conductance-vs-v4}.
%\begin{figure}[ht]
%\includegraphics[width=6cm]{conductance-vs-v4.eps}
%\caption{The voltage dependence of the amplitude of
%Aharonov-Bohm oscillations in  the conductance, for equal arm lengths
%and Coulomb interactions at $\beta^{\prime}=20$, $5$ and $2$.} 
%\label{conductance-vs-v4}
%\end{figure}

\subsection{Interferometer with arms of different lengths}

\begin{figure}
\includegraphics[width=8cm]{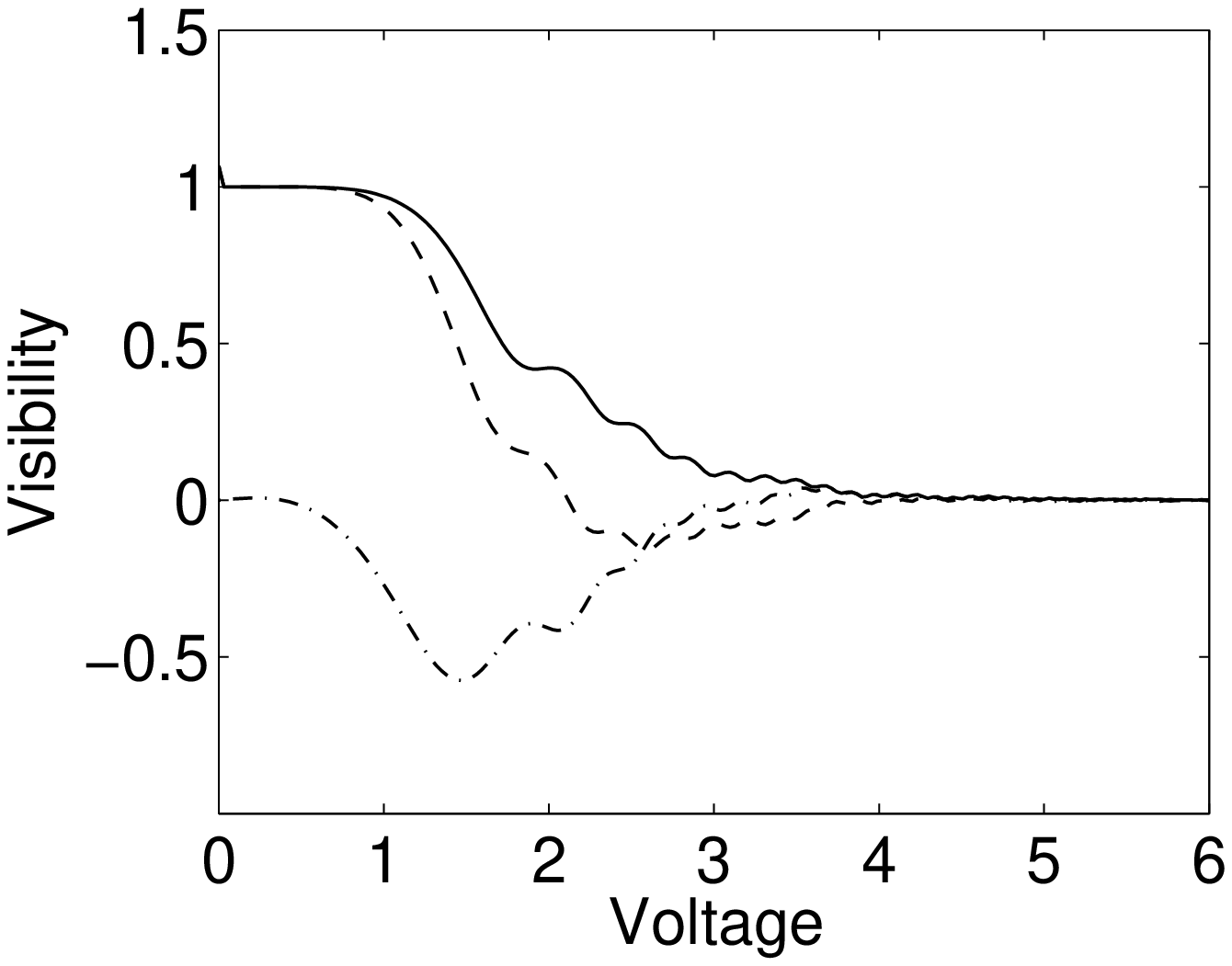}
\includegraphics[width=8cm]{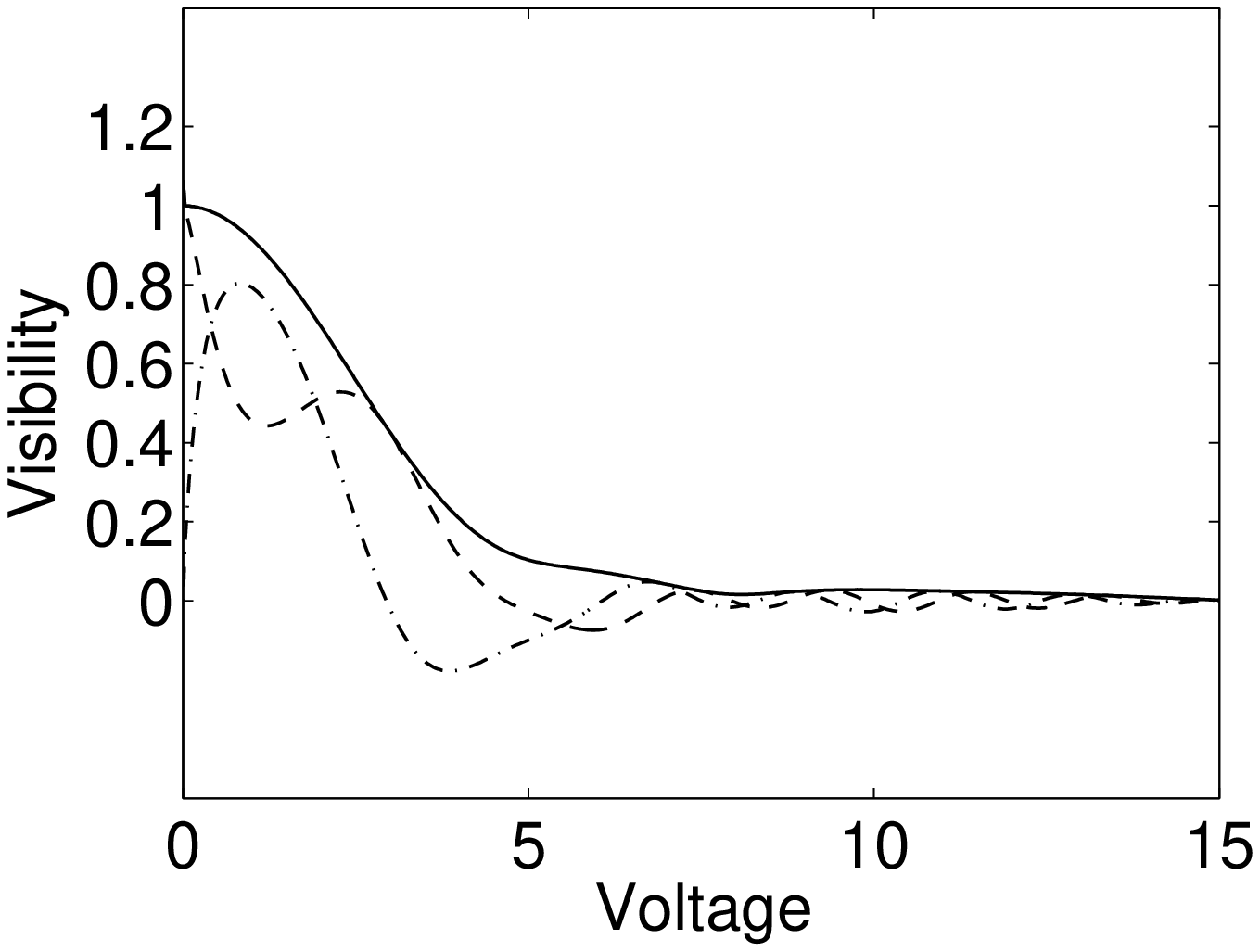}
\caption{The voltage dependence of visibility
for an interferometer having arms of different lengths, at zero temperature.
Upper panel: short-range interactions; lower panel: Coulomb
interactions.
Full lines: ${\cal V}_{\rm rel}(V^{\prime},\beta^{\prime})$; dashed
lines and dot-dashed lines: contributions from $\sigma_1(V)$ and 
$\sigma_2(V)$, respectively.} 
\label{sigma-vs-v-different-arms}
\end{figure}

For an interferometer with different arm lengths, 
the oscillatory part of the conductance is proportional to
\begin{equation}
\int_{-\infty}^{\infty}dt \exp(-ieVt/\hbar)t {\rm Im}[
g(d_1,t+d_1/v) g(d_2, t+d_1/v)]\,.\nonumber
\label{osc-g}
\end{equation}
Since $g(x,t)$ away from its peaks is asymptotically pure imaginary, the only
contributions to ${\rm Im}[g(d_1,t+d_1/v) g(d_2, t+d_1/v)]$
are from the regions of $t$ close to the peak of one or other
correlation function. 
We introduce $\ell_1$ and
$\ell_2$, obtained from $\ell$ by setting $x=d_1$ and $x=d_2$,
respectively. Similarly, we define $t_{\varphi 1}$ and $t_{\varphi
  2}$.
The peak in $g(d_1,t+d_1/v)$ occurs near $t=0$.
Close to this peak we can write $t =\tau t_{\varphi 1}$ 
and approximate
\begin{eqnarray}
g(d_1,t+d_1/v) &\approx&
\frac{1}{2\pi \ell_1}\exp(U(\beta^{\prime},\tau)
+ i W(\tau))
\nonumber \\
g(d_2, t+d_1/v)&\approx&
\frac{i \pi k_{B} T /\hbar v }{2 \pi \sinh([d_2 - d_1]\pi
  k_{B} T/\hbar v)}\,\,.\nonumber
\end{eqnarray}
The peak in the other correlation function, $g(d_2, t+d_1/v)$, is
near $t=(d_2 - d_1)/v$. Close to this peak, we can write
$t=(d_2 - d_1)/v + \tau t_{\varphi 2}$
and approximate
\begin{eqnarray}
g(d_1,t+d_1/v) &\approx&
-\frac{i \pi k_{B} T /\hbar v }{2 \pi \sinh([d_2 - d_1]\pi
  k_{B} T/\hbar v)}
\nonumber \\
g(d_2, t+d_1/v)&\approx&
\frac{1}{2\pi \ell_2} \exp(U(\beta^{\prime},\tau)
+ i W(\tau))\,.\nonumber
\end{eqnarray}
The combination $t {\rm Im}[
g(d_1,t+d_1/v) g(d_2, t+d_1/v)]$ is larger near the second peak
than
the first, by a factor $(d_2 - d_1)/\ell_1$,
because $t$ is larger in this case. When evaluating this contribution,
we need to introduce a scaled voltage $V^{\prime}{=} (eVt_{\varphi 2}/\hbar)$ 
and a scaled length difference 
$\lambda = (d_2 - d_1)/\ell_2$. 
Then the leading contribution to the integrand in Eq.~(\ref{osc-g}) 
for $d_1$, $d_2$ large 
is
\begin{eqnarray}
&&\exp(-ieVt/\hbar)t {\rm Im}[
g(d_1,t+d_1/v) g(d_2, t+d_1/v)]
\nonumber \\
&=&\exp(-i\lambda V^{\prime}) \exp(-i V^{\prime} \tau)
\frac{d_1 - d_2}{v}\nonumber\\
&&\times
\frac{\pi k_{B} T /\hbar v }{(2 \pi)^2 \sinh([d_2 - d_1]\pi
  k_{B} T/\hbar v)}\frac{1}{\ell_2}\nonumber\\
&&\times
\exp(U(\beta^{\prime},\tau))
\cos(V(\tau))\,.
%&=&
%-\exp(-i\lambda V^{\prime}) \exp(-i V^{\prime} \tau)
%\frac{1}{(2\pi)^2 v \ell_2}
%\exp(U(\beta^{\prime},\tau))
%\cos(W(\tau))\nonumber
\label{unequal}
\end{eqnarray}
The factors on the right side of Eq.~(\ref{unequal})
from the two correlation functions lead to suppression of 
conductance oscillations by two different mechanisms: thermal smearing, governed
by the parameter $k_{\rm B} T (d_2 - d_1)/\hbar v$, and interaction
effects,
governed by $k_{\rm B} T t_{\varphi 2}/\hbar$. Unless $d_2 - d_1$ is
parametrically smaller than $d_2$, thermal smearing dominates and
behaviour at non-zero temperature is obtained simply by multiplying
the zero-temperature results by the factor 
$[\pi k_{B} T /\hbar v ]/[(2 \pi)^2 \sinh([d_2 - d_1]\pi
  k_{B} T/\hbar v)]$.
%
%The voltage-dependence of the oscillatory contribution to the conductance
%is hence given at all temperatures for an interferometer with
%unequal arm lengths by the Fourier transform of
%$\exp(U(\tau))\cos(V(\tau))$.
%This function is illustrated in Fig.~\ref{FTarg-different-arms}.
%\begin{figure}
%\includegraphics[width=6cm]{FTarg-different-arms.eps}
%\caption{The function $\exp(U(0,\tau))\cos(V(\tau))$ with short range interactions: its Fourier
%  transform gives the voltage-dependence of the amplitude
%of conductance oscillations.} 
%\label{FTarg-different-arms}
%\end{figure}
%
Results for conductance as a function of voltage at zero
temperature are shown in Fig.~\ref{sigma-vs-v-different-arms}.
%The same quantity for a system with Coulomb interactions is shown in
%
%Turning to a system with Coulomb interactions, the real part of the
%Green function is shown in Fig.~\ref{FTarg-different-arms-log}, and
%the voltage-dependence of conductance oscillations is shown in
%Fig.~\ref{sigma-vs-v-different-arms-log}.
%\begin{figure}[ht]
%\includegraphics[width=6cm]{FTarg-different-arms-log.eps}
%\caption{The function $\exp(U(0,\tau)) \cos(V(\tau))$, calculated
%for Coulomb interactions: its
%Fourier transform gives the voltage dependence of the amplitude of
%Aharonov-Bohm oscillations in  the conductance, for an interferometer
%with arms of different length} 
%\label{FTarg-different-arms-log}
%\end{figure}
%\begin{figure}
%\includegraphics[width=6cm]{sigma-vs-v-different-arms-log.eps}
%\caption{The voltage dependence of the amplitude of
%Aharonov-Bohm oscillations in  the conductance
%for arms of different lengths with Coulomb interactions at zero temperature.
%Full line: $\sigma(V)$; dashed line: $\sigma_1(V)$; 
%dot-dashed line:  $\sigma_2(V)$.} 
%\label{sigma-vs-v-different-arms-log}
%\end{figure}
%\end{widetext}

%\subsection{Noise}
%
%Should say something about visibility of AB oscillations in noise
%power. Perhaps include one or two graphs.

\section{Discussion}\label{discussion}

In summary, we have shown using a microscopic treatment of
electron-electron interactions how, at
finite bias or temperature, these interactions 
influence Aharonov-Bohm oscillations
in conductance and noise power for a Mach-Zehnder interferometer.
The visibility of oscillations in both quantities
at weak tunneling is determined by the correlation function 
for electrons in a quantum Hall edge state, and we have
analysed the form of this correlation function for systems
with short range interactions or with unscreened Coulomb interactions.
Measurement of the visibility of oscillations as a function of bias 
in an interferometer with arms of equal length should provide a means
to determine this correlation function. 

Visibility is suppressed both at high bias and at
high temperature. The voltage scale for bias dependence of visibility
is set by the time scale $t_\varphi$, given in Eqns.~(\ref{t-phi-sr})
and (\ref{t-phi-coulomb}).
Suppression of visibility with temperature arises both
from dephasing by interactions, also on a scale set by $t_\varphi$,
and --- in an interferometer with arms of different lengths --- from
thermal smearing, a single-particle effect. A similar combination
of dephasing and thermal smearing contributions is well known in the
temperature dependence of mesoscopic conductance
fluctuations.\cite{altshuler}
For the interferometer, thermal smearing dominates over dephasing
unless the difference in arm lengths is small.

A discussion of numerical values for dephasing lengths is made
difficult by uncertainty in the value of the edge velocity $v$.
An experimental study of edge magnetoplasmons yields a velcoity of
about $2.5 \times 10^4{\rm ms^{-1}}$ in $\rm GaAs$ with a field of 5T at filling factor
$\nu=1$. Alternatively,
an upper limit is $v \sim \omega_c l_{\rm B}$, where $\omega_c$ is the
cyclotron frequency; taking the effective mass of electrons in $\rm
GaAs$ and a field of $4$T, this yields $v\sim 1.4 \times 10^5\, {\rm
  ms}^{-1}$. The resulting value of the thermal length at a
temperature of $100 \,{\rm mK}$ is $L_{\rm T}  = 10 \,\mu{\rm m}$.
At this temperature the dephasing length due either to short range
interactions or to curvature in the electron disperison relation
far exceeds experimental sample dimensions. To see this, consider for
definiteness an edge state confined by a metallic gate, and suppose
that Coulomb interactions are screened by this gate so that
the interaction range is set by the depletion length: $b \sim D \sim
200 \,{\rm nm}$. Then, from Eq.~(\ref{ell-phi-sr}) with $n=3$, $\ell_\varphi \sim\, 25\, {\rm
  mm}$, while $\ell^{\rm curv}_\varphi \sim 1 \,{\rm mm}$. By contrast, the
dephasing length due to unscreened Coulomb interactions, being only
logarithmically larger than $L_{\rm T}$, is comparable to sample
dimensions, and may be smaller if the true value of $v$ is smaller. 
Consistency with experiment therefore requires use of unscreened Coulomb 
interactions in the theory we have developed.
We note that sample edges in the experiments of
Refs.~\onlinecite{heiblum1} - \onlinecite{heiblum3} are defined mainly
by etching, so that interactions within the edge states are likely to
be poorly screened.

Earlier work has treated dephasing more phenomenologically, via
a fictitious dephasing voltage probe\cite{Chung} or a fluctuating
external potential.\cite{Seelig01,Marquardt04,Forster,Marquardt06}
Our results from a microscopic approach complement this earlier work
in several important ways. First, we connect the voltage and
temperature scales for suppression of Aharonov-Bohm oscillations with
parameters characterising the system at a microscopic level: the edge
velocity and the interaction range. Second, we demonstrate that the
temperature-dependence of the dephasing length depends very much on
the nature of electron-electron interactions within the system:
with realistic parameter choices, interactions that are short range because of external screening
generate negligible dephasing, while unscreened Coulomb interactions
are much more effective in reducing coherence. Third, we show that the
functional form for the dependence of visibility with temperature or
voltage varies significantly according to the model chosen for
interactions, and so provides a useful window through which details
of scattering processes can be viewed.

Turning to a comparison of our results with the experimental 
observations of Refs.~\onlinecite{heiblum1}
-- \onlinecite{heiblum3}, while our calculations capture
the central phenomenon of suppression of visibility with either
increasing temperature or increasing bias,\cite{heiblum1}
there are also features that we cannot account for, and others
that we cannot make contact with, because our calculations
are at leading order in tunneling amplitudes. In more detail, 
some successes and limitations are as follows. In
Ref.~\onlinecite{heiblum1}, suppression of visibility of oscillations
in the differential conductance is observed, with the same energy
scale entering both temperature and voltage dependence. Our
calculations reproduce this provided the interferometer is taken to have equal length
arms. In the same paper, it is suggested that noise power retains
coherent features at a bias large enough to suppress visibility in
differential conductance. The observations concerned were made with
$|\tau_b|^2 =  0.5$, which places them far outside the regime in which
our results can be applied; nevertheless, we note that such behaviour
is in contrast to our results, in the sense that a common scale enters
the visibility of oscillations in conductance and noise. 
The idea that there is a common scale for both quantities is in
qualitative agreement with a more recent
set of data \cite{heiblum3}. In
Ref.~\onlinecite{heiblum2}, very striking lobes were reported in 
visibility as a function of bias at low temperature. It is noteworthy
that the dependence we find of visibility on bias
is not in all cases simply a monotonic decrease, and, for Coulomb
interactions and equal arm lengths (Fig.~\ref{b=2-conductance}), shows pronounced
peaks and minima. Within our calculations, these oscillations appear
because visibility is given by a Fourier transform of the correlation
function representing propagation of an injected electron along one
arm of the interferometer, and because the injected electron is
represented by a pulse with a rather sharp trailing edge.
It is tempting to relate these calculated oscillations to the observed lobes.
However, a further key observation\cite{heiblum2} is that a change
in length for one of the interferometer arms leads to a reduction in
the amplitude of these lobes in visibility, without change in their
period, while our calculations give no such lobes in an interferometer
with arms of substantially different length (see
Fig.~\ref{sigma-vs-v-different-arms}).
For future work, there would be great interest in studying more
realistic models of interactions, perhaps taking better account of the
geometry of the sample, to see whether visibility lobes are a robust
consequence.

\acknowledgments
We thank M. Heiblum and D. G. Polyakov for valuable discussions. This work was supported by EPSRC Grant Nos.
GR/R83712/01 and GR/S45492/01, by the U.S.-Israel BSF and by the ISF
of the Israel Academy of Sciences. 
It was completed while JTC and MYV were visitors at KITP Santa
Barbara, supported by NSF Grant No. PHY99-07949.

%%%%%%%%%%%%%%%%%%%%%%%%%%%%%%%%%%%%%%%%%%%%%%%%%%%%%%%%%%%%%%%%%%%%%%%%%%

\end{document}